\documentclass[fleqn,usenatbib]{mnras}

\usepackage{newtxtext,newtxmath}

\usepackage[T1]{fontenc}

\usepackage{chemformula} 
\usepackage[T1]{fontenc} 
\usepackage{multirow}
\usepackage{bigdelim}
\usepackage{longtable}
\usepackage{graphicx}
\usepackage{times}
\usepackage{amsmath}
\usepackage[utf8]{inputenc}
\usepackage[T1]{fontenc}

\title{FAUST VI. VLA 1623--2417 B: a new laboratory for astrochemistry around protostars on 50 au scale}

\author[C. Codella et al.]{
C. Codella,$^{1,2}$\thanks{E-mail: claudio.codella@inaf.it}
A. L\'{o}pez-Sepulcre,$^{2,3}$
S. Ohashi,$^{4}$
C. J. Chandler,$^{5}$
M. De Simone,$^{2,1}$
\newauthor
L. Podio,$^{1}$
C. Ceccarelli,$^{2}$
N. Sakai,$^{4}$
F. Alves,$^{6}$
A. Dur\'an,$^{7}$
D. Fedele,$^{1}$
L. Loinard,$^{7,8}$
\newauthor
S. Mercimek,$^{1,9}$
N. Murillo,$^{4}$
Y. Zhang,$^{4}$
E. Bianchi,$^{2,1}$ 
M. Bouvier,$^{2}$ 
G. Busquet,$^{10}$ 
\newauthor
P. Caselli,$^{11}$ 
F. Dulieu,$^{12}$ 
S. Feng,$^{13}$ 
T. Hanawa,$^{14}$ 
D. Johnstone,$^{15,16}$ 
B. Lefloch,$^{2}$ 
\newauthor
L. T. Maud,$^{17}$ 
G. Moellenbrock,$^{5}$ 
Y. Oya,$^{18,19}$ 
B. Svoboda,$^{5}$ 
and S. Yamamoto$^{18,19}$ 
\\
\\
$^{1}$INAF, Osservatorio Astrofisico di Arcetri, Largo E. Fermi 5, I-50125, Firenze, Italy\\
$^{2}$Univ. Grenoble Alpes, CNRS, IPAG, 38000 Grenoble, France\\
$^{3,2}$Institut de Radioastronomie Millim\'{e}trique, 38406 Saint-Martin d’H\`{e}res, France\\
$^{4}$RIKEN Cluster for Pioneering Research, 2-1, Hirosawa, Wako-shi, Saitama 351-0198, Japan\\
$^{5}$National Radio Astronomy Observatory, PO Box O, Socorro, NM 87801, USA\\
$^{6}$Max-Planck-Institut für extraterrestrische Physik (MPE), Gießenbachstr. 1, D-85741 Garching, Germany\\
$^{7}$Instituto de Radioastronomía y Astrofísica , Universidad Nacional Autónoma de México, A.P. 3-72 (Xangari), 8701, Morelia, Mexico\\
$^{8}$Instituto de Astronomía, Universidad Nacional Autónoma de México, Ciudad Universitaria, A.P. 70-264, Cuidad de México 04510, Mexico\\
$^{9}$Università degli Studi di Firenze, Dipartimento di Fisica e Astronomia, via G. Sansone 1, 50019 Sesto Fiorentino, Italy\\
$^{10}$ Departament de Física Qu\`antica i Astrofísica,  Universitat de Barcelona (IEEC-UB), c/ Martí i Franq\`es 1, 08028, Barcelona, Spain\\
$^{11}$ Center for Astrochemical Studies, Max-Planck-Institut für extraterrestrische Physik (MPE), Gießenbachstr. 1, D-85741 Garching, Germany\\
$^{12}$ CY Cergy Paris Université, Sorbonne Université, Observatoire de Paris, PSL University, CNRS, LERMA, F-95000, Cergy, France\\
$^{13}$ Department of Astronomy, Xiamen University, Zengcu'oan West Road, Xiamen, 361005 China\\
$^{14}$ Center for Frontier Science, Chiba University, 1-33 Yayoi-cho, Inage-ku, Chiba 263-8522, Japan\\
$^{15}$ Department of Physics and Astronomy, University of Victoria, 3800 Finnerty Road, Elliot Building Victoria, BC, V8P 5C2, Canada\\
$^{16}$ NRC Herzberg Astronomy and Astrophysics 5071 West Saanich Road, Victoria, BC, V9E 2E7, Canada\\
$^{17}$European Southern Observatory, Karl-Schwarzschild Str. 2, 85748 Garching bei München, Germany\\
$^{18}$ Department of Astronomy, The University of Tokyo, 7-3-1 Hongo, Bunkyo-ku, Tokyo 113-0033, Japan\\
$^{19}$ Research Center for the Early Universe, The University of Tokyo, 7-3-1 Hongo, Bunkyo-ku, Tokyo 113-0033, Japan\\
}

\date{Accepted XXX. Received YYY; in original form ZZZ}
\date{Accepted XXX. Received YYY; in original form ZZZ}

\pubyear{2022}

\begin{document}
\label{firstpage}
\pagerange{\pageref{firstpage}--\pageref{lastpage}}
\maketitle

\begin{abstract}
The ALMA interferometer, with its unprecedented combination of high-sensitivity and high-angular resolution, allows for (sub-)mm wavelength mapping of protostellar systems at Solar System scales.
Astrochemistry has benefited from imaging interstellar complex organic molecules in these jet-disk systems.
Here we report the first detection of methanol (CH$_3$OH) and methyl formate (HCOOCH$_3$) emission towards the triple protostellar system VLA1623--2417 A1+A2+B, obtained in the context of the ALMA Large Program FAUST.
Compact methanol emission is detected in lines from $E_{\rm u}$ = 45 K up to 61 K and 537 K towards components A1 and B, respectively.
LVG analysis of the CH$_3$OH lines towards VLA1623--2417 B
indicates a size of 0$\farcs$11--0$\farcs$34 (14-45 au),
a column density $N$(CH$_{3}$OH) = 10$^{16}$--10$^{17}$ cm$^{-2}$,
kinetic temperature $\geq$ 170 K, and volume density $\geq$ 10$^{8}$ cm$^{-3}$.
An LTE approach is used for VLA1623--2417 A1, given the limited $E_{\rm u}$ range, and yields $T_{\rm rot}$ $\leq$ 135 K.
The methanol emission around both VLA1623--2417 A1 and B shows velocity gradients along the main axis of each disk.
Although the axial geometry of the two disks is similar, the observed velocity gradients are reversed. The CH$_3$OH spectra from B shows two broad (4--5 km s$^{-1}$) peaks, which are red- and blue-shifted by $\sim$ 6--7 km s$^{-1}$ from the systemic velocity. Assuming a chemically enriched ring within the accretion disk, close to the centrifugal barrier, its radius is calculated to be 33 au. The methanol spectra towards A1 are somewhat narrower ($\sim$ 4 km s$^{-1}$), implying a radius of 12--24 au.

\end{abstract}


\section{Introduction} \label{sec:intro}

The Sun-like star-forming process transforms dust and gas within a molecular cloud into a star surrounded by its planetary system \citep[e.g.][and references therein]{Andre2000,Frank2014}. During each evolutionary phase, matter evolves chemically increasing its complexity \citep[e.g.][and references therein]{Ceccarelli2007,Herbst2009,Caselli2012}. The earliest protostellar phases are represented by Class 0 and I objects (10$^4$-10$^5$ yr) and characterised by three major components: (i) an infalling and rotating envelope, (ii) an accretion disk, rotating along the protostellar equatorial plane, and feeding the star, and (iii) a fast ($\sim$100 km s$^{-1}$) jet and slower disk wind shedding angular momentum to allow the system to continue accreting mass. In addition, the inner 100\,au of the protostellar region is associated with a temperature $>$100\,K, which make dust mantles sublimate and in turn enrich the gas mixture. Additionally, heating via shocks is expected where the infalling material  meets the disk, close to the centrifugal barrier \citep{Stahler1994}. As a consequence, chemically enriched rotating rings are predicted and have recently been observed in several objects, starting from the prototypical L1527 \citep{Sakai2014a,Sakai2014b,Sakai2017,Oya2016}

One of the breakthrough lessons provided by the ALMA (Atacama Large Millimeter Array) interferometer\footnote{https://www.almaobservatory.org} is that rings and gaps exist in protoplanetary disks around stars younger than 1 Myr \citep[e.g.][]{Sheehan2017,Fedele2018,Segura2020}. This indicates that the  process of planet formation starts earlier than commonly thought. These findings in turn highlight the importance of studying the chemical content at the Class 0/I stages, especially imaging interstellar Complex Organic Molecules (iCOMs; species with at least  6 atoms, as e.g. CH$_3$OH), considered the first step towards a true prebiotic chemistry \citep{Herbst2009,Ceccarelli2017}. This is one of the goals of  the ALMA Large Program (LP) FAUST\footnote{http://faust-alma.riken.jp}  (Fifty AU STudy of the chemistry in  the disk/envelope system of Solar-like protostars), focused on astrochemistry of protostars imaged at the Solar System spatial scale. A full description of the FAUST project is presented by \citet{Codella2021}.  Key questions investigated are: do all Sun-like analogs pass through a hot-corino phase, and/or are they associated with chemical enrichment in the external regions (rings) of the accretion disk? FAUST allows also the opportunity to sample  the chemistry of star-forming regions in the southern hemisphere, opening new protostellar laboratories in which to investigate the composition of gas in the regions where planets are going to form.

In this paper we provide the first survey of iCOMs emission from the southern protostellar cluster VLA1623--2417, reporting detection of CH$_3$OH (methanol) and HCOOCH$_3$ (methyl formate).

\subsection{The VLA1623--2417 multiple protostars} \label{sec:target}

One of the best laboratories in which to study the chemical composition around multiple protostars is VLA 1623--2417  (hereafter VLA 1623), located in Ophiucus A  at a distance of 131$\pm$1 pc \citep{Gagne2018}. VLA 1623 is a well known multiple system with two sources labelled A and B (separated by about 1$\arcsec$, $\sim$130\,au), previously traced from cm to submm \citep[e.g.][and references therein]{Andre1990,Andre1993,Leous1991,Looney2000,Ward2011,Murillo2018L,Murillo2018}. VLA1623 A has been considered one of the prototypical Class 0 objects \citep{Andre1993,Murillo2013disk}. \citet{Harris2018}, however, recently showed that source A is actually a binary system composed of two objects, A1 and A2, separated by less than 30\,au, and surrounded by a circumbinary disk well detected at 0.9\,mm. The nature of source B, on the other hand, is still controversial \citep{Murillo2013,Murillo2013disk,Murillo2018L,Murillo2018}: (i) it is associated with water masers \citep{Furuya2003} and it has a Spectral Energy Distribution similar to VLA1623 A \citep{Murillo2018}, and (ii) it lies outside of the A1+A2 circumbinary disk \citep[see also][]{Hsieh2020}.  In addition, \citet{Hsieh2020} suggest the occurrence of SO detected accretion flows  on scales larger than 100\,au moving towards VLA1623 B.

About 1200\,au west of the VLA1623 A1+A2+B triple system, a further object, labelled W, has been revealed \citep[e.g.][and references therein]{Harris2018}. It has been proposed that VLA1623 W is a shocked cloudet, part of the large scale outflow driven by A \citep[e.g.][]{Hara2021}. However, \citet{Murillo2013}, analysing the spectral energy distribution, revealed it as a Class I object \citep[see also][]{Harris2018}. Large-scale outflowing material (e.g.\ in CO and H$_2$) originating near the VLA1623 A and B multiple system has been detected along a NW-SE direction \citep[e.g.][and references therein]{Andre1990,Caratti2006}, though the number of flows as well as the driving sources are yet to be fully understood. \citet{Santangelo2015} imaged a fast jet from VLA1623 B while \citet{Hsieh2020} analysed VLA1623 in several molecular tracers (CO isotoplogues, SO, DCO$^+$) with ALMA and reported the occurrence of two cavities at the same position but moving at different velocities, proposing that VLA1623 A and VLA1623 B are driving two molecular  outflows on the plane of the sky and on top of each other.  More recently, \citet{Hara2021} traced VLA1623 in CO with ALMA observing two outflows along the projected NW-SE direction, but in this case proposing A1 and A2 as the driving sources. Finally, \citet{Ohashi2022}, as part of ALMA-FAUST, sampled the 50\,au spatial scale using CS, CCH, and HCO$^+$ and finding: (i) a unique, wide, rotating, and low-velocity cavity (with a PA of $\sim$125$\degr$, i.e.\ NW-SE) opened by A1; (ii) a large-scale ($\sim$2000\,au) envelope as well as a circumbinary disk around A1 and A2, also rotating with the same sense of the outflow cavity, and (iii) CS emission tracing the disk around VLA1623 B, which is rotating in the opposite direction with respect to the other components of the system.

Wrapping up, the dust and the molecular content of the VLA1623 cluster previously has been extensively investigated; the missing piece of the puzzle is to understand the molecular complexity around the 4 protostars within the cluster via an iCOM survey. 

\section{Observations} \label{sec:obs}

The VLA1623 multiple system was observed on 2018 December, 2019 April, and 2020 March with ALMA Band 6 (FAUST Large Program 2018.1.01205.L), using different configurations from 40 to 49 antennas. We observed two frequency ranges (i) 214.0--219.0\,GHz and 229.0--234.0\,GHz (Setup 1), and (ii) 242.5--247.5\,GHz and 257.2--262.5\,GHz (Setup 2). Both Setups 1 and 2 were observed over 12 spectral windows with a bandwidth/frequency resolution of 59\,MHz/122\,kHz (82\,km\,s$^{-1}$/0.17--0.20 km\,s$^{-1}$) and one with a bandwidth of 1.9\,GHz (2640--2798 km\,s$^{-1}$). For the latter window, the frequency resolution was 0.5\,MHz (0.72\,km\,s$^{-1}$) for Setup 1, and 1\,MHz (1.39\,km\,s$^{-1}$) for Setup 2. The baselines were between 15\,m ($B_{\rm min}$) and 969\,m ($B_{\rm max}$). The maximum recoverable scale ($\theta_{\rm MRS}$\,$\sim$\,$0.6\,\lambda\,B_{\rm min}^{-1}$) is $\sim\,$40$\arcsec$--45$\arcsec$.

The observations were centered at $\alpha_{\rm J2000}$ = 16$^{\rm h}$\,26$^{\rm m}$\,26$\fs$392, $\delta_{\rm J2000}$ = --24$\degr$\,24$\arcmin$\,30$\farcs$178. The flux was calibrated using the quasars J1427-4206, J1517-2422, and J1626-2951, reaching an absolute flux calibration uncertainty of 10\%. The data were self-calibrated 
using line-free continuum channels. The ALMA calibration pipeline within \textsc{CASA 5.6.1} \citep{McMullin2007} was used and we included an additional calibration routine to correct for $T_{\rm sys}$ issues and spectral data normalization\footnote{https://help.almascience.org/kb/articles/what-errors-could-originate-from-the-correlator-spectral-normalization-and-tsys-calibration; Moellenbrock et al. (in preparation)}. The resulting continuum-subtracted line-cube were cleaned with a Briggs robust parameter of 1. The typical synthesized beams are 0$\farcs$45$\times$0$\farcs$36 (PA=+96$^{\circ}$), for Setup 1, and 0$\farcs$46$\times$0$\farcs$43 (PA=--80$^{\circ}$), for Setup 2.  The typical r.m.s.\ noise is $\sim$\,1--2\,mJy beam$^{-1}$. Self-calibration improved the dynamic range of the continuum images by factors between 3 to 10, depending on the dataset and configuration. The final r.m.s. noise is as expected for the integration time and bandwidth. The data analysis was performed using the \textsc{IRAM-GILDAS}\footnote{http://www.iram.fr/IRAMFR/GILDAS} package.


\begin{table*}
\caption{Spectral Properties and observed velocity integrated intensities of the CH$_3$OH and HCOOCH$_3$ lines observed towards the position of the VLA1623--2417 A and B sources (see Figure 1 and Section 3). Upper limits on CH$_3$CHO, NH$_2$CHO, and CH$_3$OCH$_3$ lines are also reported.}
\label{Tab:lines}
\begin{tabular}{lccccccc}
\hline
Transition & $\nu^{\rm a}$ & E$_{\rm u}^{\rm a}$ & g$_{\rm u}^{\rm a}$ & Log$_{\rm 10}$(A$_{\rm ul}$/s$^{-1})^{\rm a}$ &
$S\mu^{\rm 2a}$ & 
\multicolumn{2}{c}{$F$} \\ 
 & (GHz) & (K) & & & (D$^2$) &  \multicolumn{2}{c}{(mJy beam$^{-1}$ km s$^{-1}$)}   \\
  & &  & & & & A & B \\
\hline
CH$_{\rm 3}$OH 4$_{\rm 2,3}$--3$_{\rm 1,2}$ E & 218440.063 & 45 & 36 & --4.3 & 13.9 & 16.3(2.3) & 66.5(3.8) \\
CH$_{\rm 3}$OH 10$_{\rm 3,7}$--11$_{\rm 2,9}$ E & 232945.797 & 190 & 84 & --4.7 & 12.1 & $\leq$ 9.9 & 36.7(3.2) \\
CH$_{\rm 3}$OH 18$_{\rm 3,15}$--17$_{\rm 4,14}$ A & 233795.666 & 447 & 148 & --4.7 & 21.9 & $\leq$ 9.8 & $\leq$ 15.0 \\
CH$_{\rm 3}$OH 4$_{\rm 2,3}$--5$_{\rm 1,4}$ A & 234683.370 & 61 & 36 & --4.7 & 4.5 & 19.9(2.3) & 32.8(6.2)$^{\rm b}$ \\
CH$_{\rm 3}$OH 5$_{\rm 4,2}$--6$_{\rm 3,3}$ E & 234689.519 & 123 & 44 & --5.2 & 1.9 & $\leq$ 9.9 &  11.1(3.3)$^{\rm b}$ \\
CH$_{\rm 3}$OH 5$_{\rm 1,4}$--4$_{\rm 1,3}$ A & 243915.788 & 50 & 44 & --4.2 & 15.5 & 32.0(2.3) & 77.2(3.9) \\
CH$_{\rm 3}$OH 20$_{\rm 3,17}$--20$_{\rm 2,18}$ A & 246074.605 & 537 & 164 & --4.1 & 73.7 & $\leq$ 6.2 & 34.4(2.9)$^{\rm c}$ \\
CH$_{\rm 3}$OH 19$_{\rm 3,16}$--19$_{\rm 2,17}$ A & 246873.301 & 490 & 156 & --4.1 & 73.7 & $\leq$ 7.3 & 35.5(3.2) \\
CH$_{\rm 3}$OH 16$_{\rm 2,15}$--15$_{\rm 3,13}$ E & 247161.950 & 338 & 132 & --4.6 & 19.3 & $\leq$ 8.3 & 39.4(3.2) \\
CH$_{\rm 3}$OH 4$_{\rm 2,2}$--5$_{\rm 1,5}$ A & 247228.587 & 61 & 36 & --4.7 & 4.3 & 16.8(2.8) & 37.2(4.8) \\ 
CH$_{\rm 3}$OH 18$_{\rm 3,15}$--18$_{\rm 2,16}$ A & 247610.918 & 447 & 148 & --4.1 & 69.4 & $\leq$ 9.5 & 35.2(3.8) \\
CH$_{\rm 3}$OH 12$_{\rm 6,7}$--13$_{\rm 5,8}$ E & 261704.409 & 360 & 100 & --4.8 & 8.5 & $\leq$ 12.3 & $\leq$ 20.4 \\
\hline
CH$_{\rm 3}$CHO 11$_{\rm 1,10}$--10$_{\rm 1,9}$ E & 216581.930 & 65 & 46 & --3.5 & 69.0 & $\leq$ 8.0 &  $\leq$ 9.8 \\
\hline
NH$_{\rm 2}$CHO 12$_{\rm 0,12}$--11$_{\rm 0,11}$ & 247390.719 & 78 & 25 & --3.0 & 156.3 & $\leq$ 9.9 & $\leq$ 12.8 \\
NH$_{\rm 2}$CHO 12$_{\rm 2,10}$--11$_{\rm 2,9}$ & 260189.090 & 92 & 25 & --2.9 & 152.6 & $\leq$ 14.8 & $\leq$ 21.6 \\
\hline

CH$_3$OCH$_3$ 18$_{\rm 5,13}$--18$_{\rm 4,14}$ AE & 257911.036   & 191 & 74 & --4.1 & 32.5 & \multirow{4}{*}{$\leq$ 13.5$^{\rm d}$} & \multirow{4}{*}{$\leq$ 13.5$^{\rm d}$} \\
CH$_3$OCH$_3$ 18$_{\rm 5,13}$--18$_{\rm 4,14}$ EA & 257911.175   & 191 & 148 & --4.1 & 64.9 &\\
CH$_3$OCH$_3$ 18$_{\rm 5,13}$--18$_{\rm 4,14}$ EE & 257913.312 & 191 & 592 & --4.1 & 259.7 & &
\\
CH$_3$OCH$_3$ 18$_{\rm 5,13}$--18$_{\rm 4,14}$ AA & 257915.519 & 191 & 222 & --4.1 & 57.4 & \\
\\

CH$_3$OCH$_3$ 14$_{\rm 1,14}$--13$_{\rm 20,13}$ EA & 258548.819 & 93 & 116 & --3.9 & 113.2 & \multirow{4}{*}{$\leq$ 15.2$^{\rm e}$} & \multirow{4}{*}{$\leq$ 22.3$^{\rm e}$} \\
CH$_3$OCH$_3$ 14$_{\rm 1,14}$--13$_{\rm 20,13}$ AE & 258548.819 & 93 & 174 & --3.9 & 75.5 & \\
CH$_3$OCH$_3$ 14$_{\rm 1,14}$--13$_{\rm 20,13}$ EE & 258549.063 & 93 & 464 & --3.9 & 301.9 & \\
CH$_3$OCH$_3$ 14$_{\rm 1,14}$--13$_{\rm 20,13}$ AA & 258549.308  & 93 & 290 & --3.9 & 188.7 &\\
\hline
HCOOCH$_{\rm 3}$ 19$_{\rm 17,2}$--18$_{\rm 17,1}$ E &
233212.773 & 123 & 78 & --3.7 & 48.0 & \multirow{2}{*}{$\leq$ 18.2$^{\rm f}$} & \multirow{2}{*}{36.9(6.2)$^{\rm f}$} \\
HCOOCH$_{\rm 3}$ 19$_{\rm 4,16}$--18$_{\rm 4,15}$ A &
233226.788 & 123 & 78 & --3.7 & 48.0 &\\
\\
HCOOCH$_{\rm 3}$ 19$_{\rm 4,14}$--18$_{\rm 4,13}$ E &
233753.960 & 114 & 74 & --3.7 & 45.8 & $\leq$ 10.8 & $\leq$ 15.2  \\
HCOOCH$_{\rm 3}$ 19$_{\rm 4,14}$--18$_{\rm 4,14}$ A &
233777.521 & 114 & 74 & --3.7 & 45.8 & $\leq$ 10.8 & $\leq$ 13.8 \\
HCOOCH$_{\rm 3}$ 20$_{\rm 10,10}$--19$_{\rm 0,9}$ E &
246600.012 & 190 & 82 & --3.8 & 40.0 & $\leq$ 12.2 & $\leq$ 15.2 \\
\\
HCOOCH$_{\rm 3}$ 20$_{\rm 10,11}$--19$_{\rm 10,10}$ A &
246613.392 & 190 & 82 & --3.8 & 40.0 & \multirow{2}{*}{$\leq$ 12.1$^{\rm g}$} & \multirow{2}{*}{$\leq$ 15.3$^{\rm g}$} \\
HCOOCH$_{\rm 3}$ 20$_{\rm 10,10}$--19$_{\rm 10,9}$ A &
246613.392 & 190 & 82 & --3.8 & 40.0 & \\
\\
HCOOCH$_{\rm 3}$ 20$_{\rm 9,11}$--19$_{\rm 9,10}$ E &
247040.650 & 177 & 82 & --3.7 & 42.5 & \multirow{6}{*}{$\leq$ 22.8$^{\rm h}$} & \multirow{6}{*}{90.0(5.3)$^{\rm h}$} \\
HCOOCH$_{\rm 3}$ 20$_{\rm 10,11}$--19$_{\rm 10,10}$ A &
247044.146 & 140 & 86 & --3.7 & 54.0 &  \\
HCOOCH$_{\rm 3}$ 21$_{\rm 3,19}$--20$_{\rm 3,18}$ E &
247053.453 & 140 & 86 & --3.7 & 54.0 &  \\
HCOOCH$_{\rm 3}$ 20$_{\rm 9,11}$--19$_{\rm 9,10}$ E &
247057.259 & 178 & 82 & --3.7 & 42.5 &  \\
HCOOCH$_{\rm 3}$ 20$_{\rm 9,11}$--19$_{\rm 9,10}$ A &
247057.737 & 178 & 82 & --3.7 & 42.5 &  \\
HCOOCH$_{\rm 3}$ 20$_{\rm 9,12}$--19$_{\rm 9,11}$ E &
247063.662 & 177 & 82 & --3.7 & 42.5 &  \\
\\
HCOOCH$_{\rm 3}$ 21$_{\rm 7,14}$--20$_{\rm 7,13}$ E &
261715.518 & 170 & 86 & --3.6 & 48.7 & $\leq$ 16.7 & $\leq$ 21.3 \\
\hline
\end{tabular}
\\
$^{\rm a}$ Spectroscopic parameters of CH$_{\rm 3}$OH, NH$_2$CHO, and CH$_3$OCH$_3$ are from \citet{Xu1997}, \citet{Xu2008}, \citet{Moti2012}, and \citet{Endres2009}, retrieved from the CDMS database \citep{Muller2005}. For CH$_{\rm 3}$CHO and HCOOCH$_3$, we refer to data by \citet{Kleiner1996} and \citet{Ilyushin2009}, retrieved from the JPL database \citep{Pickett1998}.\\
$^{\rm b}$ Given the blending of the 5$_{\rm 4,2}$--6$_{\rm 3,3}$ E red-shifted emission with the blue-shifted peak of the CH$_{\rm 3}$OH 4$_{\rm 2,3}$--5$_{\rm 1,4}$ A profile, the measurements should be considered as lower limits.\\ 
$^{\rm c}$ All the spectra have been resampled to a velocity reolution of 1.2 km s$^{-1}$, with the exception of the weak CH$_{\rm 3}$OH(18$_{\rm 3,15}$--18$_{\rm 2,16}$) A and HCOOCH$_{\rm 3}$(19$_{\rm 17,2}$--18$_{\rm 17,1}$) A emission, smoothed to 8 km s$^{-1}$ and 6 km s$^{-1}$, respectively. The upper limits refer to the 3$\sigma$ values. CH$_3$OH: For source A, the velocity interval is --3,+7 km s$^{-1}$. For source B the methanol emission has been integrated on the --11,+13 km s$^{-1}$ interval for all the lines but the 18$_{\rm 3,15}$--18$_{\rm 2,16}$ A one, for which we adopted --16,+19 km s$^{-1}$. HCOOCH$_3$: The emission has been integrated from --22 km s$^{-1}$ to +32 km s$^{-1}$, and in the --20,+29 km s$^{-1}$ range for 19$_{\rm 17,2}$--18$_{\rm 17,1}$ E, and 21$_{\rm 3,19}$--20$_{\rm 3,18}$ E, respectively.
$^{\rm d}$ to $^{\rm h}$ Blended lines of the same species.
\end{table*}

\section{Results} \label{sec:results}

\subsection{Continuum emission}

\begin{figure*}
\begin{center}
\includegraphics[scale=0.65]{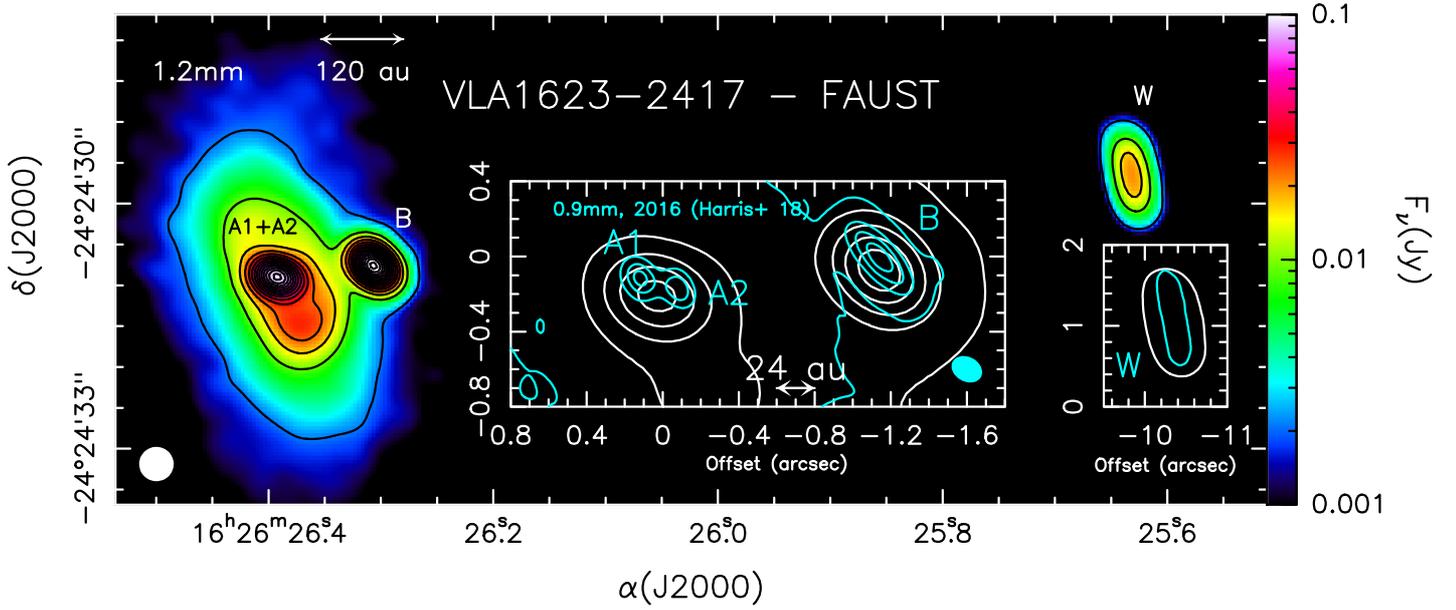}
\caption{Dust continuum emission at 1.2 mm (colour scale and black contours) from the VLA1623-2717 cluster. First contours and steps are 3$\sigma$ (2.1 mJy beam$^{-1}$) and 10$\sigma$, respectively. The synthesised beam (bottom-left corner) is 0$\farcs$39 $\times$  0$\farcs$36 (PA = 69$\degr$). The A1 and A2 protostars are not disentangled at the present angular resolutions. The B and W protostars are also labelled. The two insets compares the FAUST image (white contours, obtained in 2018) with that obtained with ALMA at 0.9 mm in 2016 (cyan contours) by \citet{Harris2018} with a higher angular resolution ($\sim$ 0$\farcs$2). For sake of clarity, for this comparison we used a step size of 30$\sigma$ for the FAUST map. Angular offsets are with respect to the phase center: $\alpha_{\rm J2000}$ = 16$^{\rm h}$ 26$^{\rm m}$ 26$\fs$392, $\delta_{\rm J2000}$ = --24$\degr$ 24$\arcmin$ 30$\farcs$178.} \label{fig:continuum}
\end{center}
\end{figure*}

Figure \ref{fig:continuum} shows the VLA 1623 region as observed in dust continuum emission at 1.2 mm. VLA1623 A is well detected, without disentangling the binary components A1 and A2 at the 0$\farcs$4 (52 au) angular resolution. The circumbinary disk around A is  well traced, in agreement with previous observations \citep[see e.g.][]{Harris2018}. Furthermore, the B and W protostars are also well detected. The J2000 coordinates of the A, B, and W protostars can be obtained from a 2D fitting: 
A: 16$^{\rm h}$\,26$^{\rm m}$\,26$\fs$392, --24$^\circ$\,24$'$\,30$\farcs$88;
B: 16$^{\rm h}$\,26$^{\rm m}$\,26$\fs$307, --24$^\circ$\,24$'$\,30$\farcs$75;
W: 16$^{\rm h}$\,26$^{\rm m}$\,25$\fs$632, --24$^\circ$\,24$'$\,29$\farcs$66.

Comparing the present continuum images with those obtained at 0.9\,mm in 2016 by \citet[][]{Harris2018} using a higher angular resolution ($\sim$0$\farcs$2; see the zoom-in insets of Fig. \ref{fig:continuum}), proper motion is suggested. For B the shift in position is $\Delta\alpha$ = +0.2\,ms and $\Delta\delta$ = --68\,mas. This shift is also in agreement with that required to align the A1+A2 positions in the map of \citet[][]{Harris2018} with the A peak in the present map. 
Further analysis of the continuum observations are beyond the scope of this paper; however, the spatial distribution of the protostars is used to determine the origin of the detected methanol emission
reported below.

\subsection{Methanol and methyl formate emission}

For the first time, the FAUST dataset allows imaging methanol (CH$_3$OH) and methyl formate (HCOOCH$_3$) line emission towards the VLA1623 protostellar system. In addition, upper limits on line emission due to acethaldeyde (CH$_3$CHO), formamide (NH$_2$CHO), and dymethyl ether (CH$_3$OCH$_3$) are derived (see Table 1).

\begin{figure*}
\begin{center}
\includegraphics[scale=0.65]{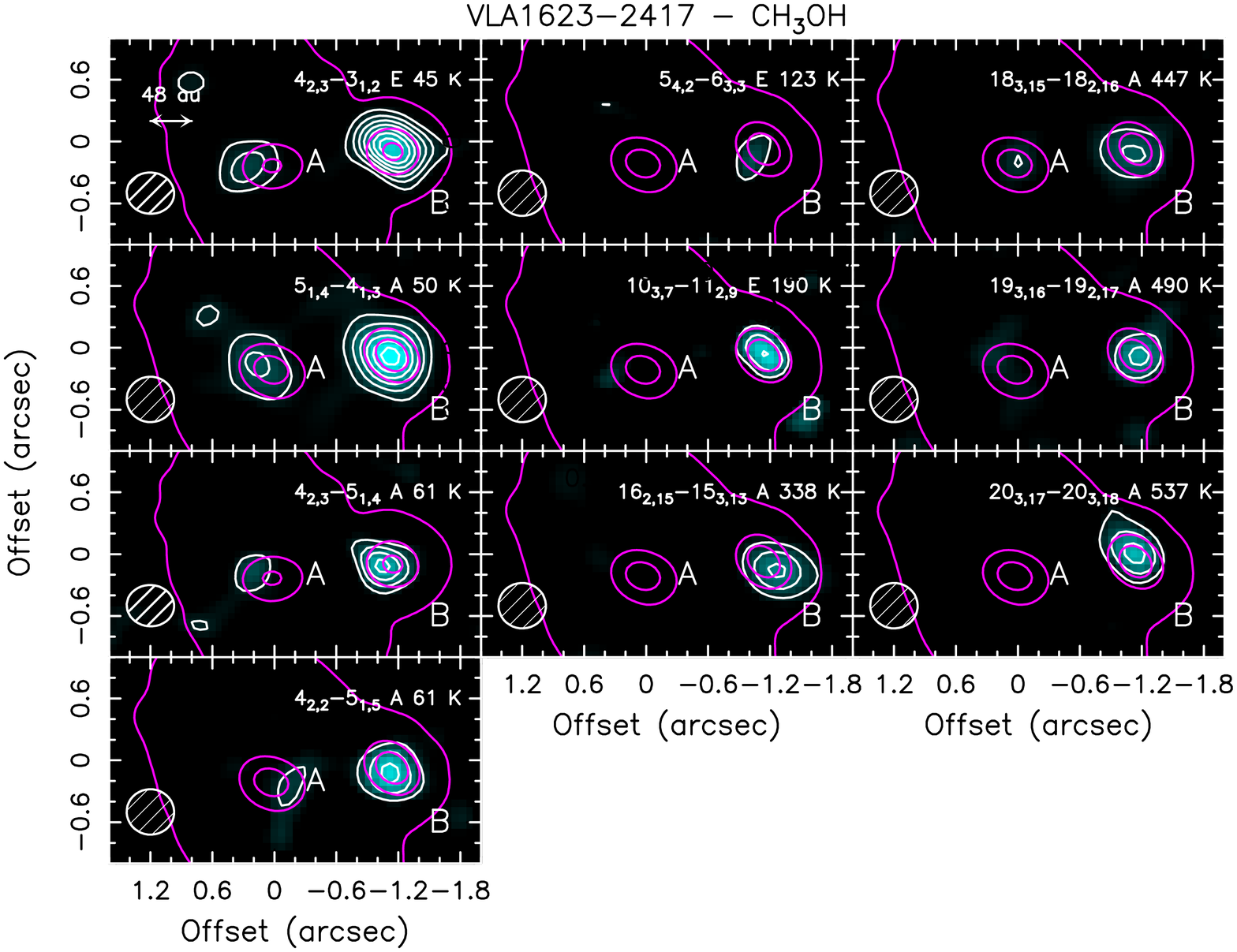}
\caption{The VLA1623 cluster as imaged using the CH$_3$OH lines (white contours) reported in Table 1. In magenta we show selected contours from the (Setup 1 and Setup 2, depending on the line) continuum emission maps, drawn to pinpoint the location of the A and B protostars. Angular offset are with respect to the phase center: $\alpha_{\rm J2000}$ = 16$^{\rm h}$ 26$^{\rm m}$ 26$\fs$392, $\delta_{\rm J2000}$ = --24$\degr$ 24$\arcmin$ 30$\farcs$178. Transitions and upper level energies are reported.  The emission has been integrated from --11\,km\,s$^{-1}$ to +14\,km\,s$^{-1}$ (for all the lines except the 20$_{\rm 3,17}$--20$_{\rm 2,18}$ A line, integrated in the --16, +24\,km\,s$^{-1}$ range, and the (blended, see Fig. \ref{fig:spectra}) 5$_{\rm 4,2}$--6$_{\rm 3,3}$ E line, integrated in the -11, +9\,km\,s$^{-1}$ range. First white contours and steps are 3$\sigma$, and 2$\sigma$, respectively.  The $\sigma$ values are 6 mJy\,km\,s$^{-1}$\,beam$^{-1}$ (5$_{\rm 1,4}$--4$_{\rm 1,3}$, 16$_{\rm 2,15}$--15$_{\rm 3,13}$, 18$_{\rm 3,15}$--18$_{\rm 2,16}$, 19$_{\rm 3,16}$--19$_{\rm 2,17}$), 4\,mJy\,km\,s$^{-1}$\,beam$^{-1}$ (4$_{\rm 2,3}$--3$_{\rm 1,2}$, 5$_{\rm 4,2}$--6$_{\rm 3,3}$, 20$_{\rm 3,17}$--20$_{\rm 2,18}$), and 5\,mJy\,km\,s$^{-1}$\,beam$^{-1}$ (4$_{\rm 2,2}$--5$_{\rm 1,5}$, 10$_{\rm 3,7}$--11$_{\rm 2,9}$). The synthesised beams (the hatched ellipse in the bottom-left corner) are 0$\farcs$45$\times$0$\farcs$36 (PA=+96$^{\circ}$), and 0$\farcs$46$\times$0$\farcs$43 (PA=--80$^{\circ}$) for Setup 1 and Setup 2, respectively.} \label{fig:lines}
\end{center}
\end{figure*}

Previously, methanol was detected through one line observed with the 12\,m Kitt Peak single dish by \citet{Lindberg2016}. Here, we observe methanol emission over eight transitions covering a large range of upper level excitation, $E_{\rm u}$, from 45\,K to 537\,K (see Table 1).  Only lines with a signal-to-noise of at least 5 are considered detected. Figure \ref{fig:lines} shows the spatial distribution of the CH$_3$OH emission lines (dust continuum emission drawn in magenta): the methanol emission peaks in two very compact regions  which are spatially unresolved at the present ($\sim$0$\farcs$4, $\sim$50 au) angular resolution. Two CH$_3$OH sources are detected: (i) one associated with the A binary protostar, and revealed only by low-$E_{\rm u}$ transitions (45--61\,K), and (ii) a second, brighter, methanol clump overlaping with the B protostar, and observed up to the $E_{\rm u}$ = 537\,K line.

\begin{figure*}
\begin{center}
\includegraphics[scale=0.62]{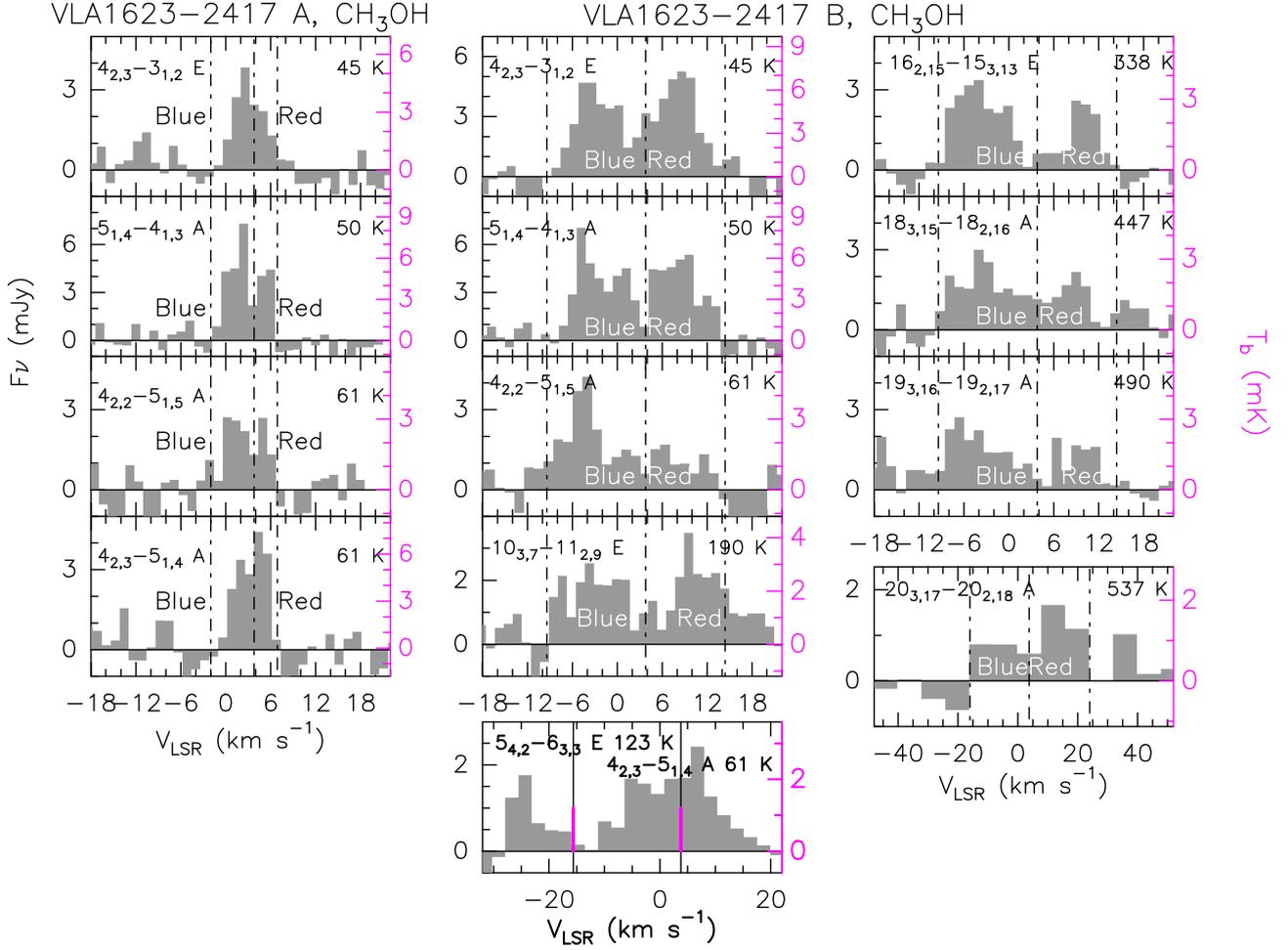}
\caption{Observed CH$_3$OH line spectra (in both F$_{\rm \nu}$ and T$_{\rm B}$ scales) at the positions of the VLA1623 A binary protostar (Left panels) and B protostar (Middle and Right panels). Transitions and upper level energies are reported. The vertical dashed lines mark: the systemic LSR velocity  \citep[+3.8 km s$^{-1}$;][]{Nara2006}, as well as the velocity ranges of the red- and blue-shifted emission used to derive velocity integrated maps (see Table 1 and Fig. \ref{fig:lines}) and maps of the blue- and red-shifted emissions (see Fig. \ref{fig:channels}). Note that the 4$_{\rm 2,3}$--5$_{\rm 1,4}$ A and 5$_{\rm 4,2}$--6$_{\rm 3,3}$ spectra towards B (smoothed to a spectral resolution of 2.4 km s$^{-1}$ to increase the S/N) are blended: the continuous black and magenta lines show the differences between their rest frequencies (in the velocity scale).} \label{fig:spectra}
\end{center}
\end{figure*}

Figure \ref{fig:spectra} reports the CH$_3$OH spectra extracted at the positions of both the A and B protostars.  The CH$_3$OH line observed with the 12\,m antenna by \citet{Lindberg2016} was very narrow (0.44\,km\,s$^{-1}$), plausibly tracing a large scale molecular envelope. Here, two different line profiles are observed. The A protostar spectra are well consistent with the systemic velocity  of +3.8\,km\,s$^{-1}$ \citep{Nara2006}, and are $\sim$4\,km\,s$^{-1}$ broad. On the other hand, the B protostar spectra show two spectrally resolved peaks (each $\sim$4--5\,km\,s$^{-1}$ broad), which are red- and blue-shifted by $\sim$6--7\,km\,s$^{-1}$ with respect to systemic. Thus, the full velocity range covered by CH$_3$OH is large, from about --12\,km\,s$^{-1}$ to +14\,km\,s$^{-1}$. Note that the 20$_{\rm 3,17}$--20$_{\rm 2,18}$ (at $E_{\rm u}$ = 537\,K) B protostar spectrum has been smoothed to 8\,km\,s$^{-1}$ in order to reach a higher signal-to-noise level; however, it stills shows a profile in agreement with the other CH$_3$OH spectra.

\begin{figure*}
\begin{center}
\includegraphics[scale=0.8]{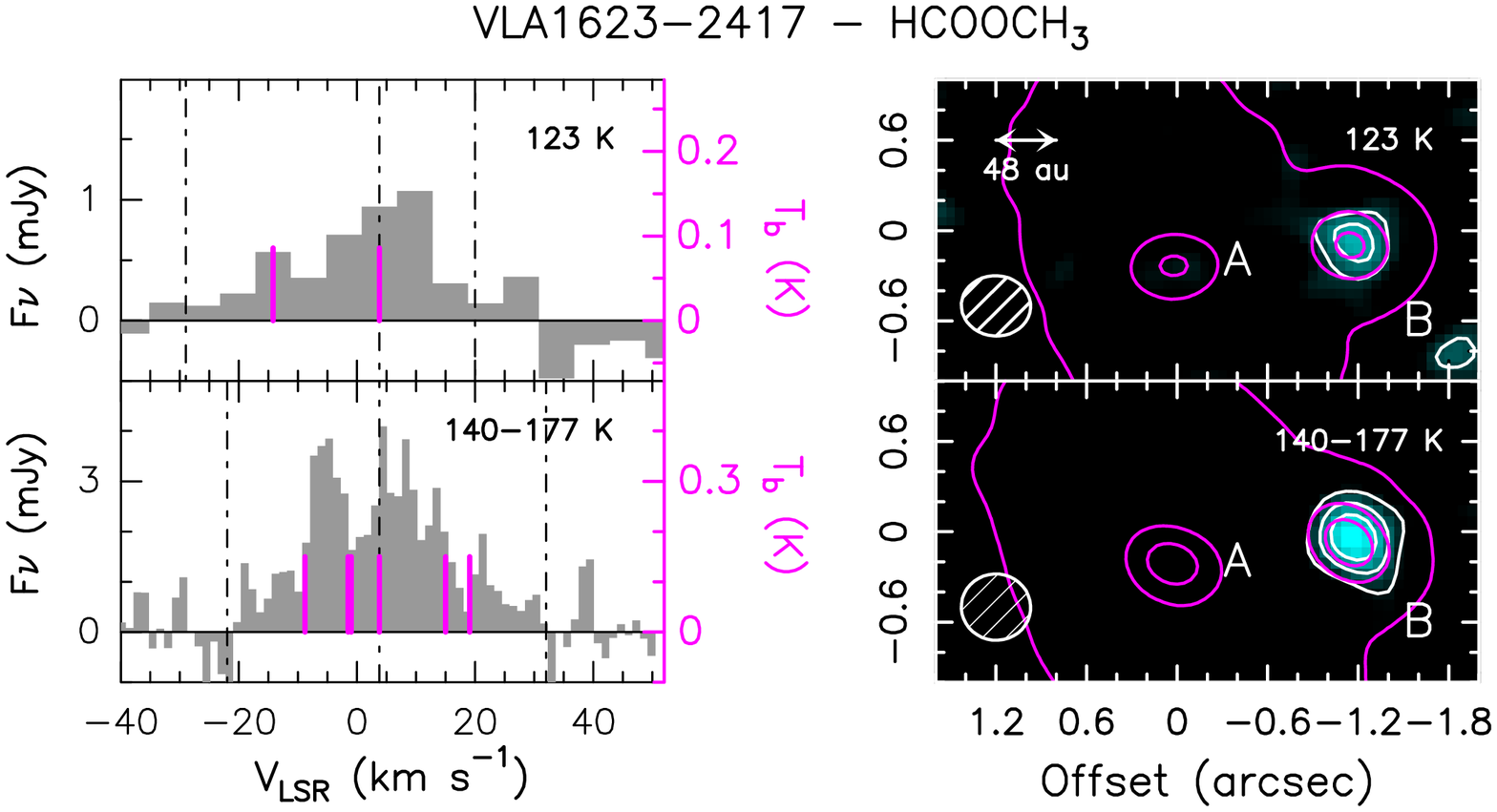}
\caption{{\it Right panels:} The VLA1623 cluster as imaged using the HCOOCH$_3$ lines (white contours) reported in Table 1. In magenta we report selected contours from the (Setup 1 and Setup 2, depending on the line) continuum emission drawn to pinpoint the A and B protostar position.  Angular offset are with respect to the phase center: $\alpha_{\rm J2000}$ = 16$^{\rm h}$ 26$^{\rm m}$ 26$\fs$392, $\delta_{\rm J2000}$ = --24$\degr$ 24$\arcmin$ 30$\farcs$178. The emission maps are due to several transitions (see Table 1), blended at the present spectral resolution. The emission has been integrated from --22\,km\,s$^{-1}$ to +32\,km\,s$^{-1}$ (covering 2 transitions with $E_{\rm u}$ = 123\,K, Upper) and in the --20, +29\,km\,s$^{-1}$ range (covering 6 transitions with $E_{\rm u}$ between 140\,K and 177\,K, Lower). First contours and steps are 3$\sigma$, and 2$\sigma$, respectively.  The $\sigma$ values are 6\,mJy\,km\,s$^{-1}$\,beam$^{-1}$ (Upper) and 8\,mJy\,km\,s$^{-1}$\,beam$^{-1}$ (Lower). The synthesised beams (the hatched ellipse in the bottom-left corner) are 0$\farcs$45$\times$0$\farcs$36 (PA=+96$^{\circ}$), and 0$\farcs$46$\times$0$\farcs$43 (PA=--80$^{\circ}$) for Setup 1 and Setup 2, respectively. {\it Left panels:} Observed HCOOCH$_3$ line spectra (in both F$_{\rm \nu}$ and T$_{\rm B}$ scales) at the positions of the B protostar. The vertical dashed line at +3.8\,km\,s$^{-1}$ marks the systemic LSR velocity \citep{Nara2006}. Other dashed lines are drawn to show the velocity range used to obtain the HCOOCH$_3$ spatial distributions. The spectra are centred at the frequency of the 19$_{\rm 17,2}$--18$_{\rm 17,1}$ E (Upper) and 21$_{\rm 3,19}$--20$_{\rm 3,18}$ E (Lower), respectively. Magenta lines indicates the shift in velocity of the other HCOOCH$_3$ lines falling in the observed spectral pattern.} \label{fig:mf}
\end{center}
\end{figure*}

HCOOCH$_3$ emission has also been revealed: (i) at 233.2\,GHz, where two transitions at $E_{\rm u}$ = 123\,K are blended, and (ii) 247.0\,GHz, where 6 transitions with $E_{\rm u}$ in the 140-177\,K range contribute to create an emission peak (see Table 1). Figure \ref{fig:mf} (Right) shows images of the HCOOCH$_3$ emission, peaking towards the B protostar. No significant emission has been observed towards the A binary protostar. The methyl formate source is spatially unresolved ($\leq$ 50\,au). The spectra extracted at the position of the B protostar are reported in Figure \ref{fig:mf} (Left). The emission covers a wide range of velocities, up to about $\pm$30\,km\,s$^{-1}$ with respect to the systemic velocity. Taking into account that the spectral patterns are associated with multiple transitions, the observed HCOOCH$_3$ profiles are in agreement with the occurrence of blue- and red-shifted peaks revealed by the CH$_3$OH lines.

\subsection{Physical properties: LTE and LVG analysis}

Given the lack of collisional coefficients for methanol at transitions with rotational number $J$ $\geq$ 16 \citep{Rabli2010}, our approach to deriving physical proporties is twofold: (i) a non-LTE (Local Thermodinamic Equilibrium) Large Velocity Gradient (LVG) model, as described by \citet{Ceccarelli2003}, for the sub-sample of CH$_3$OH transitions associated with collisional rates ($J$ $\leq$ 15), and (ii) an LTE model hypothesis using all observed CH$_3$OH lines.

\begin{figure}
\includegraphics[angle=-90,width=8cm]{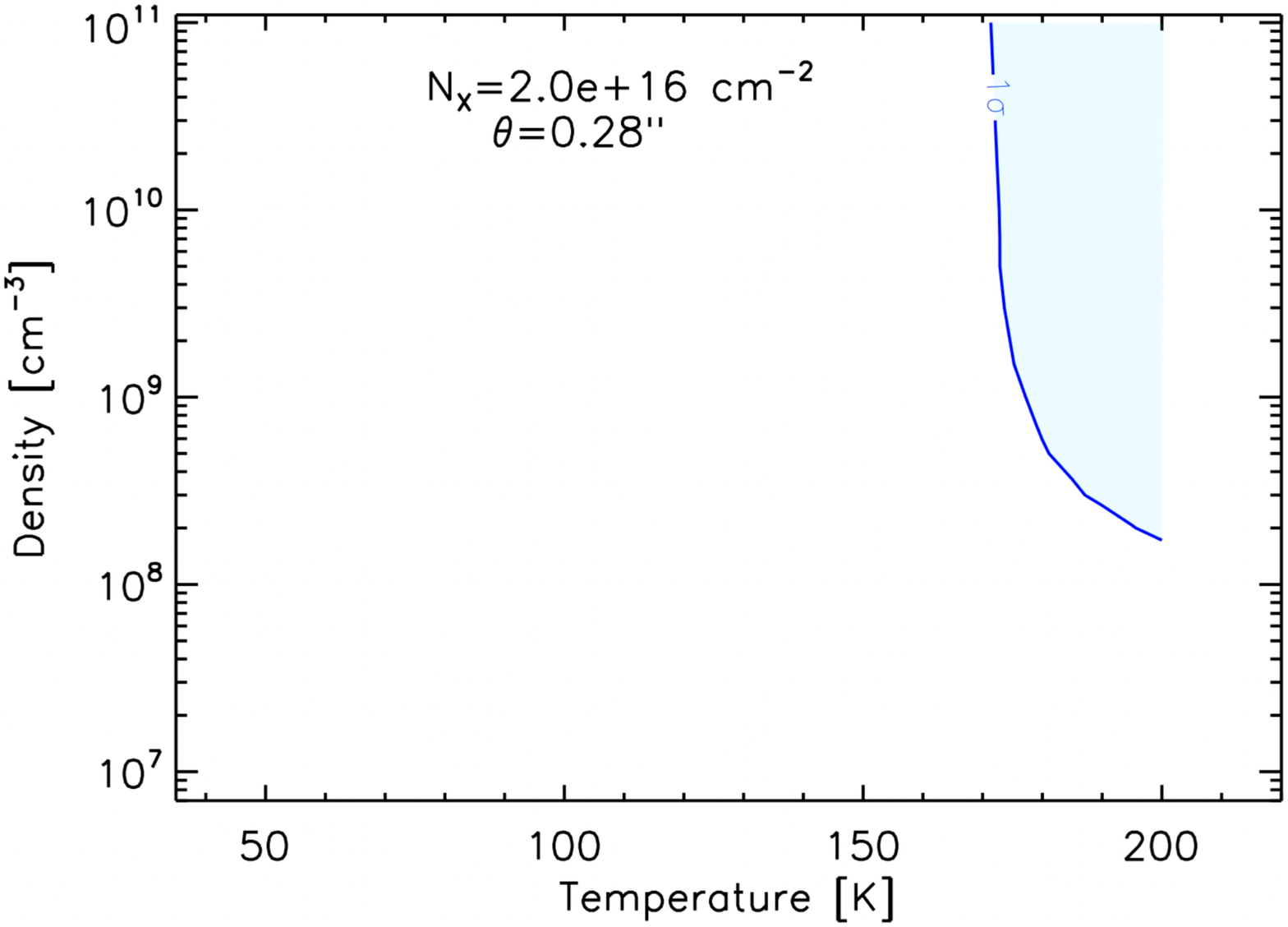} 
\includegraphics[angle=0,width=8.5cm]{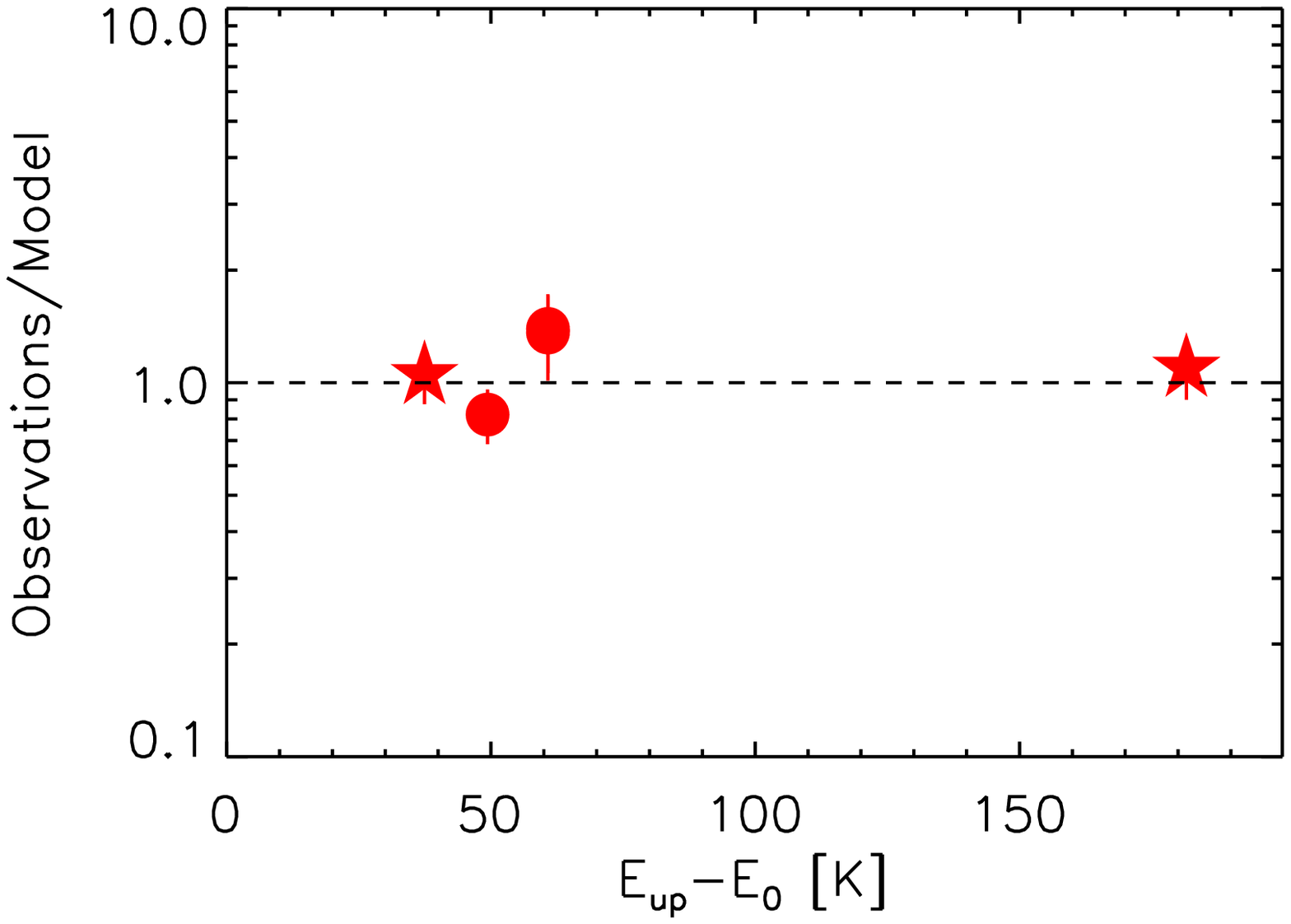}
\caption{{\it Upper panel:} Density-temperature contour plot of $\chi^{2}_r$ = 2, obtained considering the non-LTE LVG model and the observed intensity of all the A and E CH$_3$OH emission lines detected towards VLA1623 B. The best-fit case (see text) with size = 0$\farcs$28, and $N_{\rm CH_3OH}$ = 2 $\times$ 10$^{16}$\,cm$^{-2}$ is shown. The 1$\sigma$ (blue; 30\% to exceeding $\chi^{2}_r$) confidence level delimits the $n_{\rm H_{2}}$--$T_{\rm kin}$ values: $n_{\rm H_{2}}$ $\geq$\,10$^{8}$\,cm$^{-3}$, and $T_{\rm kin}$ $\geq$ 170\,K. {\it Lower panel:} Ratio between observations and model predictions of the CH$_3$OH A (circles) and E (stars) line intensity as a function of the upper level energy of the lines. The ratios are less than 2 for all the 5  methanol lines (two of them with $E_{\rm u}$ = 61\,K).}
\label{fig:lvg}
\end{figure}

For the LVG analysis toward the B protostar we use the CH$_{3}$OH-H$_{2}$ collisional coefficients computed by \citet{Rabli2010} for temperatures up to 200 K, and provided by the BASECOL database \citep{Dubernet2013}. For the computations, we use the 5 lines with $E_{\rm u}$ in the 45--190\,K range, and assume an H$_2$ ortho-to-para ratio equal to 3.  We run grids of models varying the kinetic temperature ($T_{\rm kin}$) from 50 to 200\,K, the volume density ($n_{\rm H_2}$) from 10$^{7}$ to 10$^{11}$\,cm$^{-3}$, and the methanol column density ($N_{\rm CH_3OH}$) from 10$^{15}$ to 10$^{18}$\,cm$^{-2}$.  We then simultaneously fit the measured CH$_3$OH-A and CH$_3$OH-E line intensities via comparison with those simulated by the LVG model, leaving $N$(CH$_{3}$OH), $n_{\rm H_2}$, $T_{\rm kin}$, and the emitting size $\theta$ as free parameters. The errors on the observed line intensities have been obtained by propagating the spectral r.m.s.\  with the uncertainties due to calibration (10\%). The limited number of methanol lines and the fact that 4 lines out of 5 are in the narrow  $E_{\rm u}$ = 45--61\,K range make this analysis challenging. However, several useful constraints are obtained. The lowest $\chi^{2}_r$ values ($\sim$2) are obtained for sizes between 0$\farcs$11 (14\,au) and 0$\farcs$34 (45\,au), in agreement with the observed unresolved spatial distributions. The methanol line opacities are predicted to be less than 0.1 and the A+E methanol column density $N$(CH$_{3}$OH) ranges in the 10$^{16}$--10$^{17}$\,cm$^{-2}$ interval. Figure \ref{fig:lvg} shows the $\chi^{2}_r$  contour plot in the $T_{\rm kin}$--$n_{\rm H_2}$ plane obtained for a representative case (0$\farcs$28, 2 $\times$ 10$^{16}$\,cm$^{-2}$). In addition, Figure \ref{fig:lvg} shows the ratio between observations and model predictions for the CH$_3$OH A (circles) and E (stars) line intensity as a function of the upper level energy of the lines. The 1$\sigma$ (blue; 30\% to exceeding $\chi^{2}_r$) contour delimitates the $n_{\rm H_{2}}$--$T_{\rm kin}$ degeneracy: $n_{\rm H_{2}}$ $\geq$ 10$^{8}$\,cm$^{-3}$, and  $T_{\rm kin}$ $\geq$ 170\,K (see Table \ref{Table:lvg}). These physical conditions are indeed reasonable for the inner\,50 au region around protostars as sampled by methanol \citep[e.g.,][and references therein]{Bianchi2020}.

\begin{figure}
\begin{center}
\includegraphics[scale=0.55]{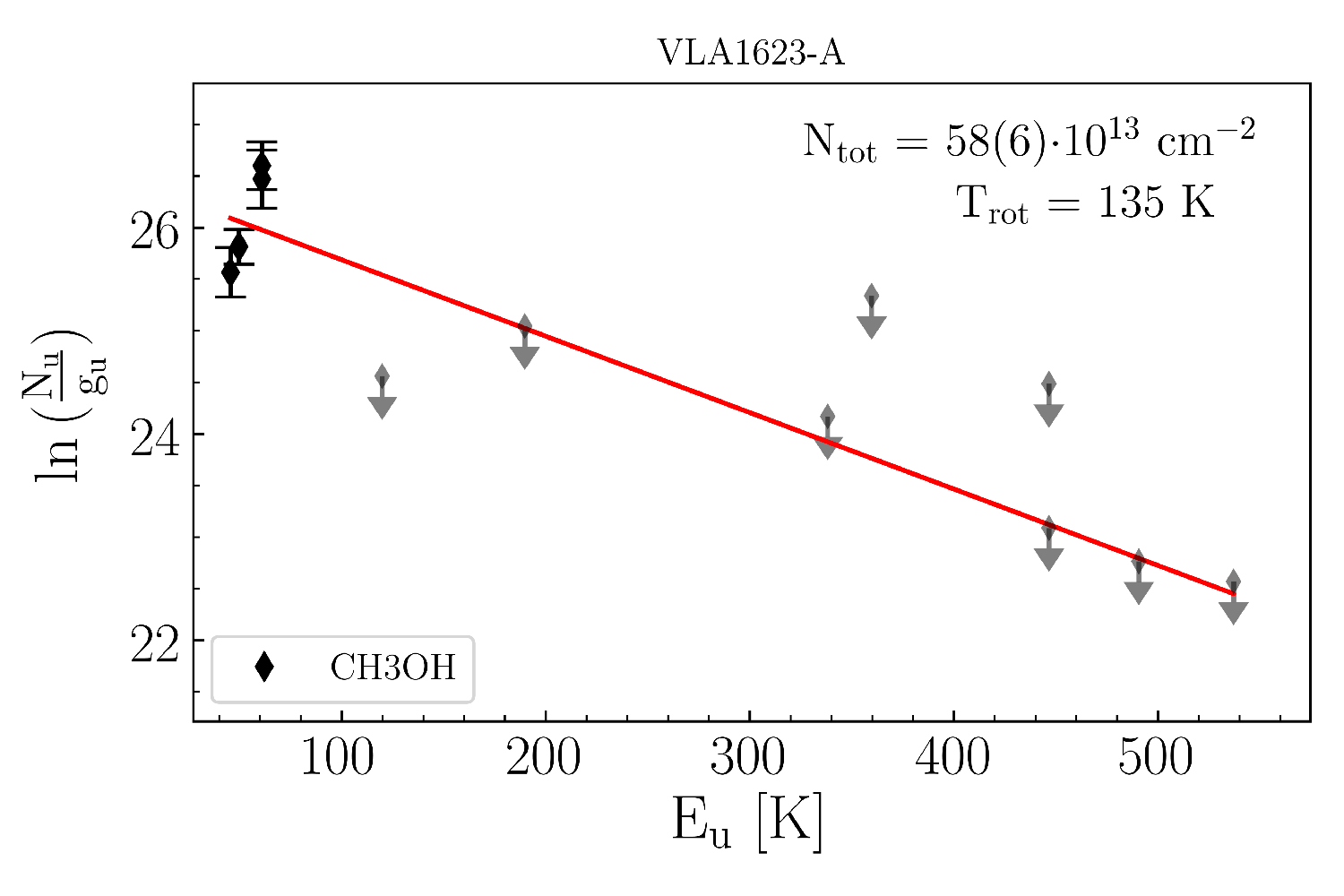} 
\includegraphics[scale=0.3,angle=-90]{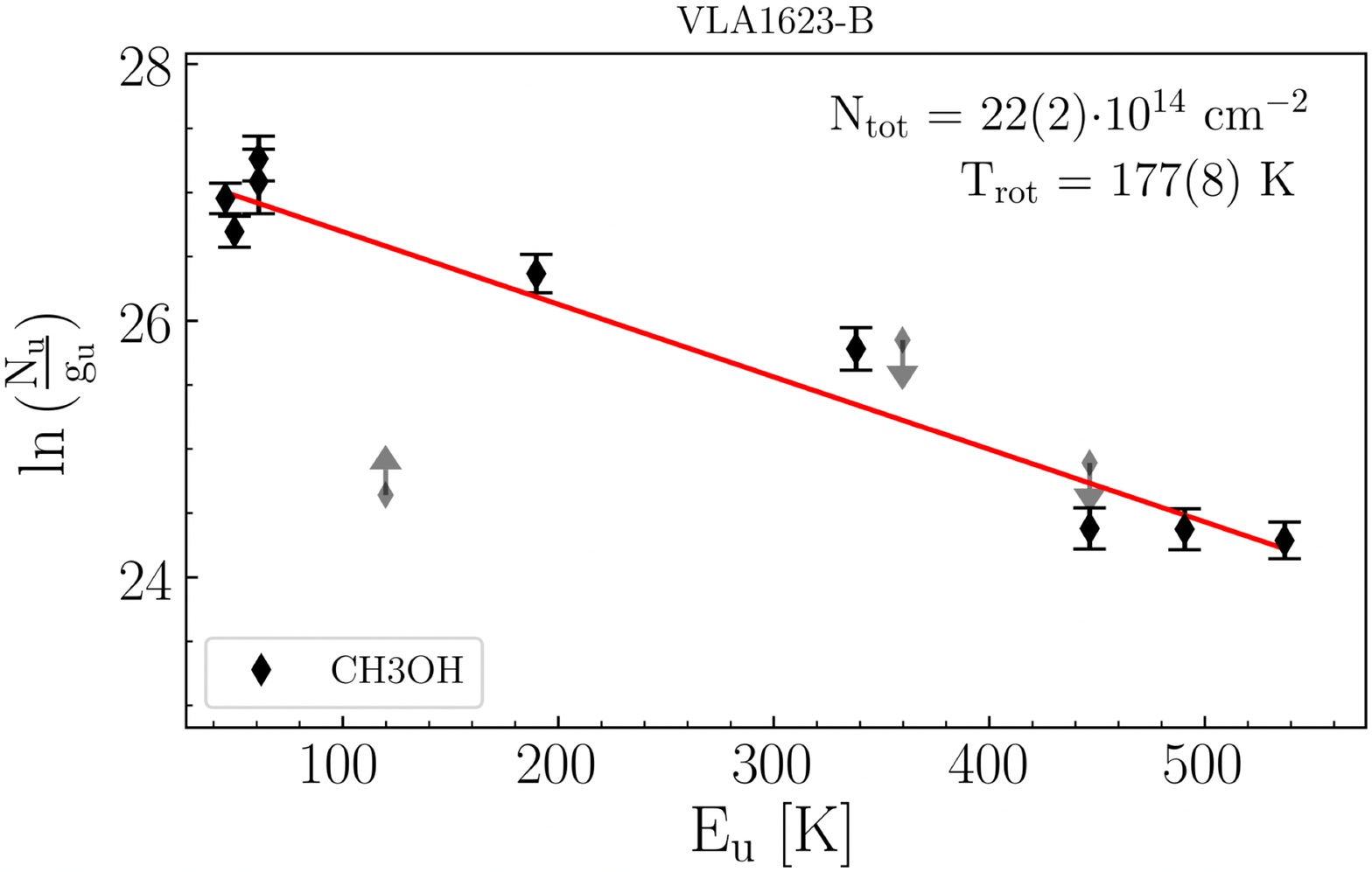}
\caption{Rotational diagrams for CH$_3$OH derived using the emission lines observed towards the continuum peaks associated with VLA1623 A and VLA1623 B (see Fig. \ref{fig:spectra}). The parameters N$_{u}$, g$_{u}$, and $E_{\rm up}$ are, respectively, the column density, the degeneracy, and the energy (with respect to the ground state of each symmetry) of the upper level.  The derived values of the rotational temperature are reported in the panels. No filling factor correction has been applied. Upper limits are reported with grey arrows. Also the lower limit derived for the 4$_{\rm 2,3}$--5$_{\rm 1,4}$ A flux towards VLA1623 B. Note that for VLA1623 A we report the fit done using the upper limits as detections: this implies that the rotation temperature of 135\,K is an upper limit (see text).} \label{fig:rd}
\end{center}
\end{figure}

Assuming an LTE population and optically thin lines (supported by the LVG analysis), we also construct rotational diagrams (RDs).  For a given molecule, the relative population distribution of all the energy levels is described by a Boltzmann temperature, that is the rotational temperature $T_{\rm rot}$. The critical densities of the CH$_3$OH lines for temperatures larger than 50\,K, when applicable \citep{Rabli2010}, are $\sim$10$^{5}$--10$^{6}$\,cm$^{-3}$. The volume densities found for the LVG analysis of methanol are indeed larger than 10$^{8}$\,cm$^{-3}$, supporting the fact that the LTE condition is satisfied. Figure \ref{fig:rd} shows the RD of CH$_3$OH, derived for both protostars A and B (see also Table 3). Upper limits are reported with grey arrows, as well as the lower limit derived for the 4$_{\rm 2,3}$--5$_{\rm 1,4}$ A flux towards VLA1623 B (see Sect. 4.2). For source B the fit provides a column density $N_{\rm tot}$ = 2.2$\pm$0.2 $\times 10^{15}$\,cm$^{-2}$ (not corrected for the filling factor), and a rotational temperature of 177$\pm$8\,K. These estimates are in very good agreement with the LVG results confirming that LTE and optically thin conditions are satisfied. In addition, the RD fit for source B ($\chi^{2}_r$ = 4) shows that even the very high-$E_{\rm u}$ (338--537\,K) lines do not require additional excitation mechanisms different from collisions, such as an IR radiation field from the protostar. For source A, the $E_{\rm u}$ range sampled by the 4 detected lines is too small to obtain a proper free fit. However, we used the upper limits on the high-excitation lines to constrain the rotational temperature. If we assume that the 3$\sigma$ upper limits on the lines with $E_{\rm u}$ larger than 300\,K are real detecions we obtain $T_{\rm rot}$ = 135\,K. The real rotational temperature has to be lower, given these upper limits for the brightness of the lines. If $T_{\rm rot}$ = 135\,K, then $N_{\rm tot}$ is $\sim$6 $\times$ 10$^{14}$\,cm$^{-2}$.  The column density decreases for lower temperature, e.g.\ down to $\sim$6 $\times$ 10$^{13}$\,cm$^{-2}$, if the rotational temperature is 50\,K.

\begin{table}
\caption{1$\sigma$ Confidence Level (range) from the Non-LTE LVG Analysis of the CH$_3$OH lines towards VLA1623 B.}
\begin{center}
\begin{tabular}{cccrr}
\hline
\multicolumn{2}{c}{Size} &        
\multicolumn{1}{c}{N$_{\rm tot}$} &
\multicolumn{1}{c}{T$_{\rm kin}$} &                                           \multicolumn{1}{c}{n$_{\rm H_2}$} \\
\multicolumn{1}{c}{(arcsec)} &
\multicolumn{1}{c}{(au)} &
\multicolumn{1}{c}{(cm$^{-2}$)} &
\multicolumn{1}{c}{(K)} &
\multicolumn{1}{c}{(cm$^{-3}$)}\\
\hline
 11--34 & 14-45 & 10$^{16}$--10$^{17}$ & $\geq$ 170 & $\geq$ 10$^{8}$ cm$^{-3}$ \\
 \hline
\end{tabular}\\
 \end{center}
\label{Table:lvg}
\end{table}

Methyl formate emission has been detected through two spectral patterns containing contributions by different transitions and, in one case, different upper level excitations. In order to take into account the multiple transitions we assume (i) the same line profiles for all transitions, and (ii) the same rotational temperature obtained from the LTE analysis of methanol. More specifically, the total column density is obtained by fitting the line profiles using the GILDAS--Weeds tool \citep{Maret2011}, finding $N_{\rm HCOOCH_3}$ = 8 $\times$ 10$^{14}$\,cm$^{-2}$ and $\leq$ 2--3 $\times$ 10$^{14}$\,cm$^{-2}$ (not corrected for filling factor effects) for sources B and A, respectively (see Table 3). Finally, transitions of acethaldeyde (CH$_3$OCH), formamide (NH$_2$CHO), and  dymethyl ether (CH$_3$OCH$_3$) fall inside the observed frequency windows, but no emission over 3$\sigma$ has been found. Table 3 reports the upper limits on the total column density derived for these species using the same methodology adopted for methyl formate: $N_{\rm CH_3CHO}$ $\leq$ 1--3 $\times 10^{14}$\,cm$^{-2}$,  $N_{\rm NH_2CHO}$ $\leq$ 0.1--5 $\times 10^{14}$\,cm$^{-2}$,  and $N_{\rm CH_3OCH_3}$ $\leq$ 4--9 $\times 10^{14}$\,cm$^{-2}$. These values are not corrected for the filling factors. If we take as representative the LVG analysis of methanol emission towards VLA1623 B, the filling factor ranges from 7 $\times$ 10$^{-2}$ to 0.41.

In summary, the compact CH$_3$OH and HCOOCH$_3$ emission, and the high rotational temperatures are consistent with thermal sublimation of the methanol molecules from icy mantles at temperatures higher than 100\,K, namely the classical definition of a hot-corino \citep{Ceccarelli2003}. This conclusion specifically applies to source B, while the origin of methanol emission in source A is less constrained. These findings are further discussed in Sect. 4.1, taking into account the observed kinematics.

\begin{table}
\caption{Results of the LTE rotational diagram analysis of the iCOMs emission observed towards VLA163 A and VLA1623 B. These values are not corrected for the filling factor derived for CH$_3$OH in VLA1623 B, which ranges from 7 $\times$ 10$^{-2}$ to 0.41 (see text).}
\begin{tabular}{lcccc}
  \hline
    \multicolumn{1}{c}{}&\multicolumn{2}{c}{VLA1623 A}&\multicolumn{2}{c}{VLA1623 B} \\
    \hline
  \multicolumn{1}{c}{Species} &\multicolumn{1}{c}{T$_{\rm rot}$} &\multicolumn{1}{c}{N$_{\rm tot}$} &\multicolumn{1}{c}{T$_{\rm rot}$} &\multicolumn{1}{c}{N$_{\rm tot}$} \\
  \multicolumn{1}{c}{} & \multicolumn{1}{c}{(K)} & \multicolumn{1}{c}{(cm$^{-2}$)} 
  & \multicolumn{1}{c}{(K)} & \multicolumn{1}{c}{(cm$^{-2}$)} \\
\hline

CH$_{3}$OH & 50--135$^a$ & 0.6--6 $\times 10^{14}$ & 177(8) & 22(2) $\times 10^{14}$ \\

HCOOCH$_3$  & 50--135$^b$ & $\leq$ 2--3 $\times 10^{14}$ & 177$^b$ & 8 $\times 10^{14}$ \\

CH$_{3}$CHO  & 50--135$^b$ & $\leq$ 1--3 $\times 10^{14}$  & 177$^b$ & $\leq$ 3 $\times 10^{14}$ \\ 

NH$_{2}$CHO & 50--135$^b$ & $\leq$ 0.3--5 $\times 10^{14}$  & 177$^b$ & $\leq$ 10 $\times 10^{13}$ \\ 

CH$_{3}$OCH$_{3}$  & 50--135$^b$ & $\leq$ 4--6 $\times 10^{14}$ & 177$^b$ & $\leq$ 9 $\times 10^{14}$ \\
  
\hline
\end{tabular}\\

$^a$ The upper limit of the range is constrained by the LTE analysis, while the lower end has been assumed.\\ 
$^b$ Assumed, as derived by the methanol LTE analysis. Upper limits refer to 3$\sigma$.
\label{Table:abundances}
\end{table}

\begin{figure*}
\begin{center}
\includegraphics[scale=0.9]{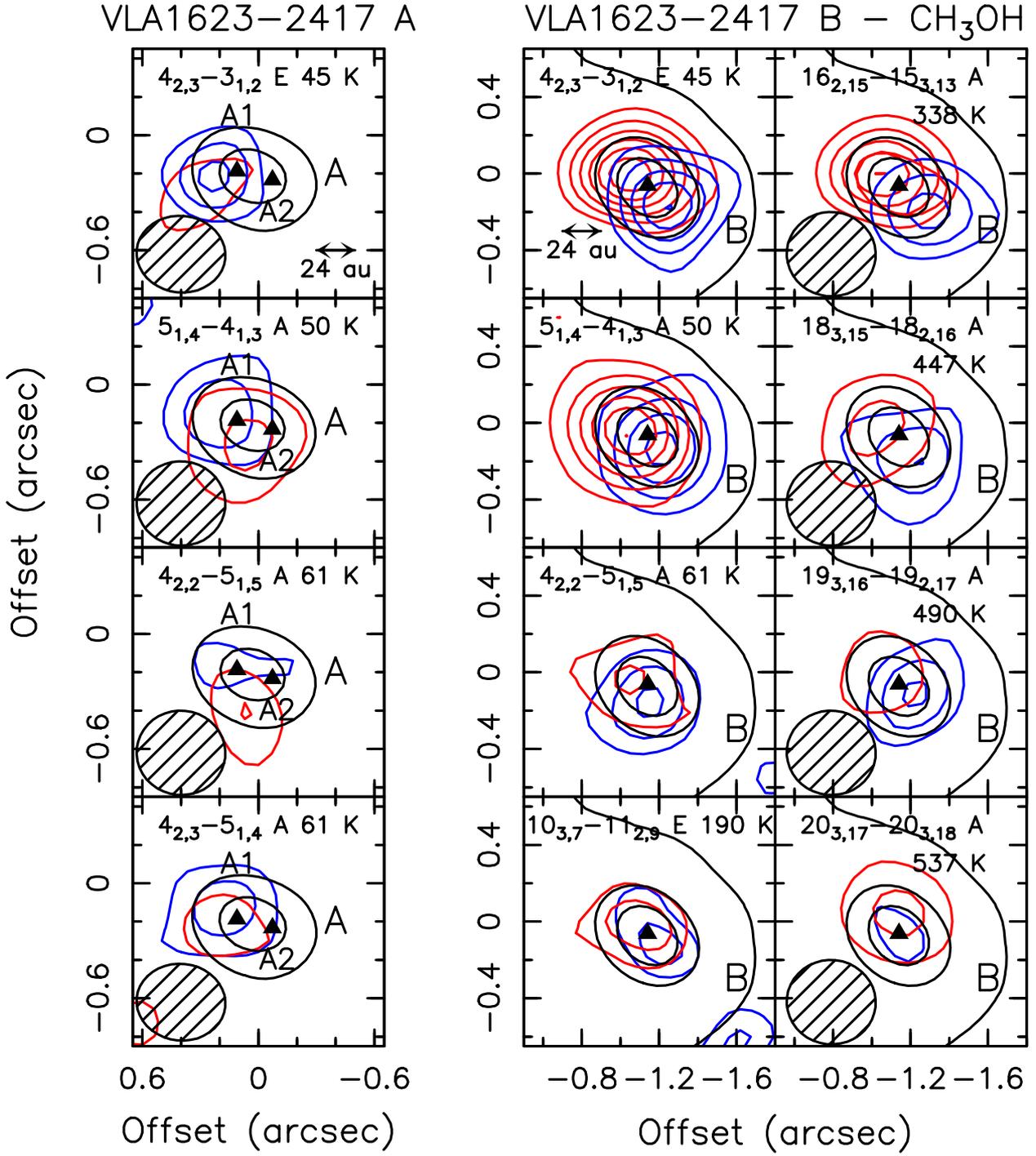}
\caption{Plot of the red- and blue-shifted CH$_3$OH emission (see Table 1) observed towards VLA1623-2717 A (Left panels), and B (Middle and Right panels) protostars. Transitions and upper level energies are reported. The systemic velocity is +3.8 km s$^{-1}$ \citep{Nara2006}. The emission has been integrated between --11\,km\,s$^{-1}$ (--2\,km\,s$^{-1}$) and  +14\,km\,s$^{-1}$ (+7\,km\,s$^{-1}$) for VLA1623-2717 B (A) for all the lines except the 20$_{\rm 3,17}$--20$_{\rm 2,18}$ A line toward VLA1623-2717 B, which is integrated over the --16,\,+24\,km\,s$^{-1}$ range. Angular offsets are with respect to the phase center. In black we report selected contours from the continuum emission (Setup 1 and Setup 2) maps, drawn to pinpoint the protostar positions.  Black triangles indicate the position of VLA1623-2717 A1, A2, and B as imaged in continuum emission by \citet{Harris2018}, obtained with a 0$\farcs$2 beam and spatially shifted taking into account the proper motion to allow for a proper comparison with the present methanol images (see text). First contour and steps are 3$\sigma$ (mJy\,km\,s$^{-1}$\,beam$^{-1}$) and 2$\sigma$, respectively.  The $\sigma$ values are 3\,mJy\,km\,s$^{-1}$\,beam$^{-1}$ for all the emission maps except: 
4$_{\rm 2,3}$--3$_{\rm 1,2}$, 
4$_{\rm 2,3}$--5$_{\rm 1,4}$,
and 4$_{\rm 2,2}$--5$_{\rm 1,5}$ 
(2\,mJy\,km\,s$^{-1}$\,beam$^{-1}$), and  
5$_{\rm 1,4}$--4$_{\rm 1,3}$,
10$_{\rm 3,7}$--11$_{\rm 2,9}$,
and 16$_{\rm 2,15}$--15$_{\rm 3,13}$ 
(4\,mJy\,km\,s$^{-1}$\,beam$^{-1}$). The synthesised beams (the hatched ellipse in the bottom-left corner) are 0$\farcs$45$\times$0$\farcs$36 (PA=+96$^{\circ}$), and 0$\farcs$46$\times$0$\farcs$43 (PA=--80$^{\circ}$) for Setup 1 and Setup 2, respectively.}
\label{fig:channels}
\end{center}
\end{figure*}

\begin{figure}
\includegraphics[scale=0.57]{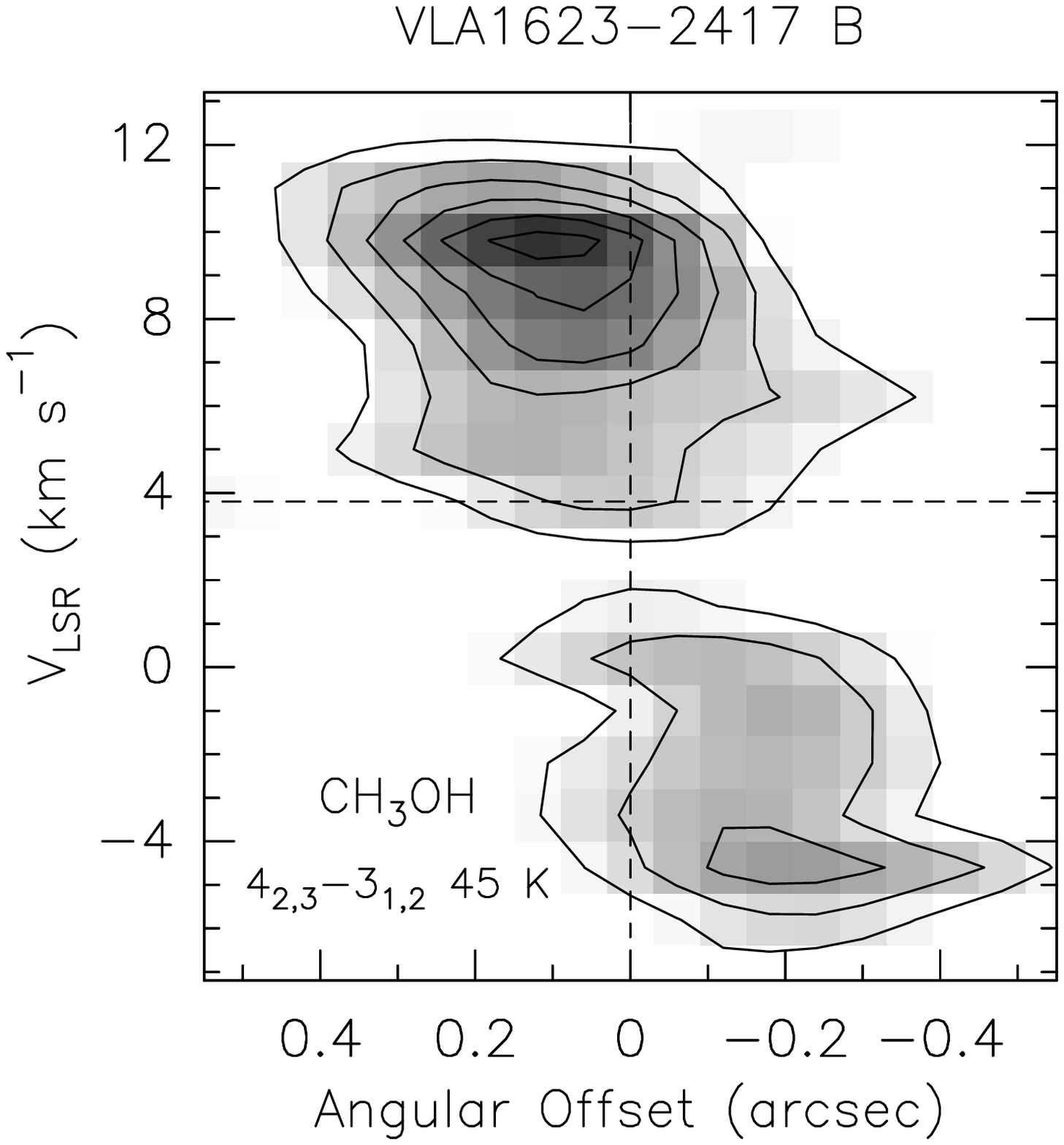}
\caption{Position-Velocity cut (beam averaged) of CH$_3$OH(4$_{\rm 2,3}$--3$_{\rm 1,2}$) E along the NE-SW disk around VLA1623 B \citep{Harris2018}. Contour levels range from 3$\sigma$ by steps of 2$\sigma$ (128 mK). Dashed
lines mark the position of the B protostar and the systemic
velocity \citep[+3.8 km s$^{-1}$;][]{Nara2006}. The angular resolution
along the disk axis is 0$\farcs$41, while the spectral resolution is 1.2 km s$^{-1}$.}
\label{fig:pv}
\end{figure}


\section{Discussion} \label{sec:discussion}

\subsection{Kinematics: on the origin of chemical enriched gas in VLA1623 A+B}

Figure \ref{fig:channels} shows the red- and blue-shifted CH$_3$OH emission observed towards the VLA1623 A (Left panels) and B (Middle and Right panels) protostars. The emission is integrated between --11\,km\,s$^{-1}$ (--2\,km\,s$^{-1}$) and  +14\,km \,s$^{-1}$ (+7\,km\,s$^{-1}$) for  VLA1623 B (A) for all the lines eacept 20$_{\rm 3,17}$--20$_{\rm 2,18}$ A, which is integrated in the --16,\,+24\,km\,s$^{-1}$ range toward VLA1623 B. At high velocities we clearly separate spatially both the red- and blue-shifted regions. Velocity gradients are revealed. More specifically, all the methanol lines around VLA1623 B shows a red-shifted peak towards NE (with respect to the continuum peak) and a blue-shifted peak towards SW. This velocity gradient is along the main axis of the disk as traced by continuum (see e.g. the high-angular resolution map by \citet{Harris2018}, reported in Fig. \ref{fig:continuum}). The same velocity gradient, with the same axis and the same velocity spread, has been recently discovered, in the FAUST context, by \citet{Ohashi2022}, using CS(5--4) emission.  

What is the origin of this chemically enriched gas? The present data must be interpreted with caution. 
Figure \ref{fig:pv} shows the Position-Velocity (PV) diagram 
of the CH$_3$OH(4$_{\rm 2,3}$--3$_{\rm 1,2}$) E emission (representative of the
line sample) for VLA1623 B. The PV has been derived along the  
direction of the disk equatorial plane as inferred from the continuum
map of \citep{Harris2018}. Figure \ref{fig:pv} reveals the rotation pattern, indicating also a lack of emission (plausibly due to absorption) at the low-velocity ($\sim$ 2--3 km s$^{-1}$, see also the corresponding spectrum in Fig. \ref{fig:spectra}) blue-shifted emission close to the protostar. A detailed modelling is hampered by the spatial resolution
(the beam size across the disk axis is 0$\farcs$41, 53 au) of the map.
One possibility is that the rotating methanol emission is tracing the inner portion of the envelope, where the temperature is high enough to thermally evaporate the dust mantles, consistent with  the classical hot-corino scenario \citep{Ceccarelli2003}. Another intriguing possibility is that CH$_3$OH is associated with the protostellar disk, more specifically with the ring-like region where the infalling-rotating envelope gas meets the accretion disk and the gas sheds angular momentum in order to continue its trip to the protostar. In this environment, low-velocity  ($\sim$1\,km\,s$^{-1}$) accretion shocks are expected with the consequent  sputtering of dust mantle products into the gas phase. The prototypical protostar where this effect has been revealed is L1527  \citep[][]{Sakai2014a,Sakai2014b}.  According to the high-angular resolution continuum measurements from \citep{Harris2018}, the VLA1623 B disk inclination derived from the ratio between the observed minor and major axes is 74$\degr$. The protostellar mass is 1.7\,$M_{\rm \sun}$ \citep[][and references therein]{Ohashi2022}, and the two CH$_3$OH peaks are located at $\pm$ 6--7\,km\,s$^{-1}$ with respect to the systemic velocity. If we assume methanol is tracing an inclined Keplerian disk, the bulk of the emission would arise at a radius of 33\,au. Noticeably, this distance is comparable with both the disk size imaged in the continuum by \citet{Harris2018} as well as the methanol-emitting size derived by the present LVG analysis (14--45\,au).

Similar instructive constraints can be derived for VLA1623 A. The maps shown in Figure \ref{fig:channels} clearly indicate that: (i) CH$_3$OH is also rotating in the NE-SW direction, but with the opposite sense compared to VLA1623 B \citep[as also noted by][using CS]{Ohashi2022}, and (ii) even though unresolved, it is possible to note that the methanol emission is shifted with respect to the continuum peak toward the west, where A1 lies. In other words, out of the coeval A1+A2 binary, only A1 appears to be associated with a hot-corino. This chemical differentiation has been indeed found in the (sub-)mm spectral window for several binaries across different star-forming regions \citep[][and references therein]{Taquet2015,Desimone2017,Bianchi2020,Belloche2020,Yang2021,Bouvier2022}. Is this a real chemical differentiation, possibly related to colder conditions, or is it a signature of dust opacity playing a major role at (sub-)mm wavelengths? The latter possibility was explored by \citet{Desimone2020} for the archetypical NGC1333-IRAS4 A1+A2 binary system, in Perseus, where according to ALMA only A2 is associated with iCOMs \citep{Lopez2017}. \citet{Desimone2020} found that, observing with the JVLA at cm-wavelengths, where dust opacity is negligible, A1 is also revealed as a hot-corino. The high dust opacity in the (sub-)mm hampers the detection of iCOMs around NGC1333-IRAS4 A1. Observations at low frequencies will be necessary to provide the final answer on the chemical richness of VLA1623 A2. 

Considering further VLA1623 A1, the spectral linewidth is narrower than for VLA1623 B. Using the continuum fit by \citet{Harris2018}, for the A1 disk we derive an inclination angle of 47$\degr$. The mass of the A1 protostar is challenging to constrain, however. From line and continuum analysis, the total A1+A2 mass ranges from 0.2\,$M_{\rm \sun}$ to 0.4 $M_{\rm \sun}$ \citep{Murillo2013disk,Harris2018,Ohashi2022}. Assuming for A1 a mass of 0.1--0.2\,$M_{\rm \sun}$, the disk inclination, and that the methanol emission is coming on average from $\pm$ 2\,km\,s$^{-1}$ with respect to the envelope velocity, we derive a radius of 12--24\,au. Again, this size in agreement with $\sim$ 14--21\,au disk derived by \citet{Harris2018} from the A1 continuum.

\subsection{Complex organics around VLA1623-2417 B}

The detection of CH$_3$OH and HCOOCH$_3$ towards VLA1623 B opens a new laboratory in which to study the chemical complexity around protostars on Solar System scales. Assuming both species are emitted from the same region, the HCOOCH$_3$/CH$_3$OH abundance ratio, derived from the column densities (not corrected for the filling factor, Table 3), is $\sim$ 0.4. This value is within the range found in the literature for hot-corinos associated with Class 0 or I objects, as imaged by interferometers. If we consider the ALMA-PILS, IRAM-CALYPSO, ALMA-FAUST, and ALMA-PEACHES, a relatively large spread is revealed, from a few 10$^{-2}$ to values close to unity \citep{Jorgensen2016,Jorgensen2018,Manigand2020,Bianchi2020,Belloche2020,Yang2021}.

Given that the VLA1623 B disk is close to edge-on \citep{Harris2018}, the present CH$_3$OH maps resemble the Orion HH212-mm protostellar disk previously observed with ALMA at the same spatial scale, i.e.\ a velocity gradient of gas enriched in iCOMs around the disk \citep[e.g.][and references therein]{Codella2018}.  HH212-mm is very bright in both dust and line emission, and a large number of iCOMs have been imaged. The HCOOCH$_3$/CH$_3$OH abundance ratio, in the HH212-mm case, is 2 $\times$ 10$^{-2}$ \citep{Lee2019}. When observed at higher spatial resolution, down to 10\,au \citep{Lee2017a,Lee2017b,Lee2017c,Lee2019}, HH212-mm discloses the region traced by iCOMs: two rotating rings at a radius of $\sim$40\,au associated with the outer surface layers of the disk. The layers lie above and below the equatorial plane by about 40\,au. In the plane, (sub-)mm emission from the dust is optically thick and no iCOMs emission is detected, plausibly due to opacity effects. As with HH212, two scenarios are possible for VLA1623 B: (i) iCOMs delimitate the accretion shock, similarly to what proposed for L1527, or (ii) they arise from portions of the flared disk illuminated directly by the protostar. VLA1623 B offers an excellent opportunity to attack this question given that ALMA can reach spatial scales of $\sim$3\,au, due to the VLA1623 system being closer ($\sim$130\,pc) to the Sun compared with Orion ($\sim$400\,pc).

\section{Summary and conclusions} \label{sec:conclusions}

The FAUST ALMA Large Program has surveyed iCOMs emission from the VLA1623-2417 protostellar cluster at 1.1.mm and 1.4mm, at the spatial scale of 50\,au. The main findings are summarised as follows:

\begin{enumerate}
    \item The spatial distribution of mm-size dust emission allows us to well detect VLA1623 A, B, and W. The binary companions A1 and A2 cannot be disentangled at the present angular resolution, but the circumbinary disk is clearly revealed. A proper motion of about 210\,mas for both the A and B objects is clearly seen by comparing the present continuum image with that obtained at 0.9\,mm in 2016 by \citet[][]{Harris2018}. 
    \item The present FAUST dataset allows us to image, for the first time, methanol (CH$_3$OH) towards VLA1623-2417, using emission lines covering a wide rage of upper level excitation, $E_{\rm u}$, from 45\,K to 537\,K. Two spatially unresolved emission peaks are detected: (i) one associated with VLA1623 A and revealed by transitions up to $E_{\rm u}$ = 61\,K, (ii) another perfectly overlapping with VLA1623 B emitting up to $E_{\rm u}$ = 537\,K;
     \item From a non-LTE LVG analysis of the CH$_3$OH emission towards VLA1623 B we obtain a size = 0$\farcs$11--0$\farcs$34 (14--45\,au), and an A+E methanol column density $N$(CH$_{3}$OH) = 10$^{16}$--10$^{17}$\,cm$^{-2}$. High kinetic temperatures and high volume densities are also required: $n_{\rm H_{2}}$ $\geq$\,10$^{8}$\,cm$^{-3}$, and $T_{\rm kin}$ $\geq$\,170\,K. No LVG analyis can be done for VLA1623 A, given that it is detected only through lines in the $E_{\rm u}$ = 45--61\,K range. An LTE RD analysis, however, provides $T_{\rm rot}$ $\leq$\,135\,K, and $N_{\rm tot}$ is $\sim$0.6--6\,$\times$\,10$^{14}$\,cm$^{-2}$. 
    \item HCOOCH$_3$ emission is imaged towards VLA1623 B. Assuming the same gas conditions derived from methanol emission, we obtain a total column density (corrected for filling factor) $N_{\rm HCOOCH_3}$ = 1-2\,$\times$ 10$^{16}$\,cm$^{-2}$. The HCOOCH$_3$/CH$_3$OH abundance ratio, derived from the column densities, is $\sim$0.4, in agreement with the range of values obtained by previous interferometric measurements towards Class 0 and I hot-corinos. Upper limits on the column densities of acethaldeyde, formamide, and dymethyl ether have been also derived.
    \item Methanol emission around VLA1623 B has a clear velocity gradient along the main axis of the disk, with a red-shifted peak towards the NE and a blue-shifted peak towards the SW. The CH$_3$OH spectra show two peaks (each $\sim$4--5\,km\,s$^{-1}$ broad), which are red- and blue-shifted by $\sim$6--7\,km\,s$^{-1}$ with respect to the systemic velocity. Assuming that CH$_3$OH traces the chemically enriched ring of the accretion disk close to the centrifugal barrier (as e.g.\ in the archetypical L1527 case) then the bulk of the emission should be emitted at a radius of 33\,au, a distance comparable with the size derived from LVG (14--45\,au). Around VLA1623 A, CH$_3$OH is also rotating in the NE-SW direction but with the opposite sense in respect to VLA1623 B, and the emission is associated with A1. Thus, out of the coeval A1+A2 binary, only A1 is associated with iCOMs and a hot-corino, according to the present ALMA data. Observations at cm-wavelengths are required to verify whether the detection of iCOMs emission towards A2 is prevented by dust opacity. The spectra in this case are relatively narrow ($\sim$4\,km\,s$^{-1}$). Assuming again a rotating ring, we derive a size of 12--24\,au.
     \end{enumerate}

To conclude, thanks to the detection of CH$_3$OH and HCOOCH$_3$, VLA1623 B  can be considered a new laboratory for studying astrochemistry around protostars. The inclination of the disk is 74$\degr$, close to edge-on. The data presented here are reminiscent of the first ALMA cycle observations toward the Orion HH212-mm protostellar disk at a spatial resolution of 50--100\,au): revealing a velocity gradient of gas enriched in iCOMs around the disk. Once imaged in iCOMs at the 10\,au scale, two rotating rings associated with the outer disk surface layers, above the optically thick equitarial plane, are observed around HH212-mm \citep[][and references therein]{Lee2019}. HH212-mm has thus become a rare but key region in which to invesigate the chemical richness of the protostellar disk and its connection to either protostellar illumination or accretion shocks . VLA1623 B offers a second such region for examination, at a distance $\sim$3 times closer than Orion.

\section*{Acknowledgements}
We thank the referee P.T.P. Ho for the fruitful comments and suggestions.
This project has received funding from: 
1) the European Research Council (ERC) under the European Union's Horizon 2020 research and innovation program, for the Project “The Dawn of Organic Chemistry” (DOC), grant agreement No 741002;
2) the PRIN-INAF 2016 The Cradle of Life - GENESIS-SKA (General Conditions in Early Planetary Systems for the rise of life with SKA); 
3) a Grant-in-Aid from Japan Society for the Promotion of Science (KAKENHI: Nos. 18H05222, 20H05844, 20H05845); 
4) the Spanish FEDER under project number ESP2017-86582-C4-1-R; 
5) DGAPA, UNAM grants IN112417 and IN112820, and CONACyT, Mexico; 
6) ANR of France under contract number ANR-16-CE31-0013; 
7) the French National Research Agency in the framework of the Investissements d’Avenir program (ANR-15-IDEX-02), through the funding of the "Origin of Life" project of the Univ. Grenoble-Alpes, 
8) the European Union’s Horizon 2020 research and innovation programs under projects “Astro-Chemistry Origins” (ACO), Grant No 811312;
9) The National Research Council Canada and an NSERC Dicovery Grant to DJ.
This paper makes use of the following ALMA data: ADS/JAO.ALMA\#2018.1.01205.L. ALMA is a partnership of ESO (representing its member states), NSF (USA) and NINS (Japan), together with NRC (Canada), MOST and ASIAA (Taiwan), and KASI (Republic of Korea), in cooperation with the Republic of Chile. The Joint ALMA Observatory is operated by ESO, AUI/NRAO and NAOJ. The National Radio Astronomy Observatory is a facility of the National Science Foundation operated under cooperative agreement by Associated Universities, Inc. \\

\section*{DATA AVAILABILITY} 
The raw data will be available on the ALMA archive at the end of the proprietary period (ADS/JAO.ALMA\#2018.1.01205.L).
%

\vspace{-0.5cm}
\bibliographystyle{mnras}
\bibliography{Mybib} 




\bsp	
\label{lastpage}
\end{document}